\documentclass[fleqn,usenatbib,useAMS]{mnras}

\DeclareRobustCommand{\VAN}[3]{#2}
\let\VANthebibliography\thebibliography
\def\thebibliography{\DeclareRobustCommand{\VAN}[3]{##3}\VANthebibliography}

\usepackage[T1]{fontenc}    

\usepackage{newtxtext,newtxmath}
\usepackage{graphicx}
\usepackage{natbib}
\usepackage{bm}
\usepackage{tabularx}
\usepackage{amsmath}
\usepackage{hyperref}
\usepackage{subcaption}
\usepackage{multirow}
\usepackage{xcolor}
\usepackage{ulem}
\usepackage{threeparttable}
\usepackage{color}

\usepackage{caption}

\newcommand{\Ds}{D_{\rm S}}
\newcommand{\Dl}{D_{\rm L}}
\newcommand{\Ml}{M_{\rm L}}

\newcommand{\thetae}{\theta_{\rm E}}

\newcommand{\pie}{\pi_{\rm E}}
\newcommand{\pirel}{\pi_{\rm rel}}
\newcommand{\te}{t_{\rm E}}
\newcommand{\tEi}{t_{{\rm E},i}}
\newcommand{\murel}{\mu_{\rm rel}}
\newcommand{\eventa}{KMT-2022-BLG-0954}
\newcommand{\eventb}{KMT-2024-BLG-0697}
\newcommand{\eventc}{MOA-2024-BLG-018}

\newcommand{\hjdPrime}{{\rm HJD}^{\prime}}

\newcommand{\orcid}[1]{}     

\DeclareGraphicsExtensions{.pdf,.png,.jpg}

\title[Three Microlensing Planet Candidates]{A Comprehensive Analysis of Three Microlensing Planet Candidates with the Planet/Binary Degeneracy}

\author[Zhang et al.]{Jiyuan Zhang$^{1}$\thanks{E-mail: zhangjy22@mails.tsinghua.edu.cn}\orcid{0000-0002-1279-0666},
Weicheng Zang$^{2}$\orcid{0000-0001-6000-3463},
Yoon-Hyun Ryu$^{3}$\orcid{0000-0001-9823-2907},
Takahiro Sumi$^{4}$,
{Andrzej Udalski}$^{5}$\orcid{0000-0001-5207-5619},
\newauthor
Shude Mao$^{6}$\orcid{0000-0001-8317-2788}
\hspace{1cm} (Leading Authors)
\newauthor
Michael D. Albrow$^{7}$\orcid{0000-0003-3316-4012},
Sun-Ju Chung$^{3,8}$\orcid{0000-0001-6285-4528},
Andrew Gould$^{9}$,
Cheongho Han$^{10}$\orcid{0000-0002-2641-9964},
Kyu-Ha Hwang$^{3}$\orcid{0000-0002-9241-4117},
\newauthor
Youn Kil Jung$^{3,8}$\orcid{0000-0002-0314-6000},
In-Gu Shin$^{6}$\orcid{0000-0002-4355-9838},
Yossi Shvartzvald$^{11}$\orcid{0000-0003-1525-5041},
Jennifer C. Yee$^{2}$\orcid{0000-0001-9481-7123},
Hongjing Yang$^{6}$\orcid{0000-0003-0626-8465},
\newauthor
Sang-Mok Cha$^{3,12}$\orcid{0000-0002-7511-2950},
Dong-Jin Kim$^{3}$,
Seung-Lee Kim$^{3}$\orcid{0000-0003-0562-5643},
Chung-Uk Lee$^{3}$\orcid{0000-0003-0043-3925},
Dong-Joo Lee$^{3}$\orcid{0009-0000-5737-0908},
\newauthor
Yongseok Lee$^{3,12}$\orcid{0000-0001-7594-8072},
Byeong-Gon Park$^{3}$\orcid{0000-0002-6982-7722},
Richard W. Pogge$^{9,13}$\orcid{0000-0003-1435-3053}
\newauthor
\hspace{1cm} (The KMTNet Collaboration)
\newauthor
Yunyi Tang$^{1}$,
Leandro de Almeida$^{14,15}$,
Dan Maoz$^{16}$,
Qiyue Qian$^{1}$\orcid{0000-0003-4625-8595},
Wei Zhu$^{1}$\orcid{0000-0003-4027-4711}
\newauthor
\hspace{1cm} (The MAP Follow-up Teams)
\newauthor
Fumio Abe$^{17}$,
Ken Bando$^{4}$,
David P. Bennett$^{18,19}$,
Aparna Bhattacharya$^{18,19}$,
Ian A. Bond$^{20}$,
\newauthor
Akihiko Fukui$^{21,22}$,
Ryusei Hamada$^{4}$,
Shunya Hamada$^{4}$,
Naoto Hamasaki$^{4}$,
Yuki Hirao$^{4}$,
\newauthor
Stela Ishitani Silva$^{18,21}$,
Naoki Koshimoto$^{4}$,
Yutaka Matsubara$^{17}$,
Shota Miyazaki$^{24}$,
Yasushi Muraki$^{17}$,
\newauthor
Tutumi Nagai$^{4}$,
Kansuke Nunota$^{4}$,
Greg Olmschenk$^{18}$,
Cl\'ement Ranc$^{25}$,
Nicholas J. Rattenbury$^{26}$,
\newauthor
Yuki Satoh$^{4}$,
Daisuke Suzuki$^{4}$,
Sean K. Terry$^{18,19}$,
Paul J. Tristram$^{27}$,
Aikaterini Vandorou$^{18,19}$,
\newauthor
Hibiki Yama$^{4}$
\newauthor
\hspace{1cm} (The MOA Collaboration)
\newauthor
Przemek Mr\'{o}z$^{5}$\orcid{0000-0001-7016-1692},
Micha{\l}~K. Szyma\'{n}ski$^{5}$\orcid{0000-0002-0548-8995},
Jan Skowron$^{5}$\orcid{0000-0002-2335-1730},
Radoslaw Poleski$^{5}$\orcid{0000-0002-9245-6368},
Igor Soszy\'{n}ski$^{5}$\orcid{0000-0002-7777-0842},
\newauthor
Pawe{\l} Pietrukowicz$^{5}$\orcid{0000-0002-2339-5899},
Szymon Koz{\l}owski$^{5}$\orcid{0000-0003-4084-880X},
Krzysztof A. Rybicki$^{5,11}$\orcid{0000-0002-9326-9329},
Patryk Iwanek$^{5}$\orcid{0000-0002-6212-7221},
\newauthor
Krzysztof Ulaczyk$^{5}$\orcid{0000-0001-6364-408X},
Marcin Wrona$^{5,28}$\orcid{0000-0002-3051-274X},
Mariusz Gromadzki$^{5}$\orcid{0000-0002-1650-1518},
Mateusz J. Mr\'{o}z$^{5}$
\newauthor
\hspace{1cm} (The OGLE Collaboration)
\\
$^{1}$Department of Astronomy, Tsinghua University, Beijing 100084, China\\
$^{2}$Center for Astrophysics | Harvard \& Smithsonian 60 Garden St., Cambridge, MA 02138, USA\\
$^{3}$Korea Astronomy and Space Science Institute, Daejon 34055, Republic of Korea\\
$^{4}$Department of Earth and Space Science, Graduate School of Science, Osaka University, Toyonaka, Osaka 560-0043, Japan\\
$^{5}$Astronomical Observatory, University of Warsaw, Al. Ujazdowskie 4, 00-478 Warszawa, Poland\\
$^{6}$Department of Astronomy, Westlake University, Hangzhou 310030, Zhejiang Province, China\\
$^{7}$University of Canterbury, Department of Physics and Astronomy, Private Bag 4800, Christchurch 8020, New Zealand\\
$^{8}$University of Science and Technology, Korea, (UST), 217 Gajeong-ro Yuseong-gu, Daejeon 34113, Republic of Korea\\
$^{9}$Department of Astronomy, Ohio State University, 140 W. 18th Ave., Columbus, OH 43210, USA\\
$^{10}$Department of Physics, Chungbuk National University, Cheongju 28644, Republic of Korea\\
$^{11}$Department of Particle Physics and Astrophysics, Weizmann Institute of Science, Rehovot 76100, Israel\\
$^{12}$School of Space Research, Kyung Hee University, Yongin, Kyeonggi 17104, Republic of Korea\\
$^{13}$Center for Cosmology and AstroParticle Physics, Ohio State University, 191 West Woodruff Ave., Columbus, OH 43210, USA\\
$^{14}$SOAR Telescope/NSF’s NOIRLab, Avda Juan Cisternas 1500, 1700000, La Serena, Chile\\
$^{15}$Laborat\'orio Nacional de Astrof\'isica, Rua Estados Unidos, 154, Itajub\'a, MG, Brazil\\
$^{16}$School of Physics and Astronomy, Tel-Aviv University, Tel-Aviv 6997801, Israel\\
$^{17}$Institute for Space-Earth Environmental Research, Nagoya University, Nagoya 464-8601, Japan\\
$^{18}$Code 667, NASA Goddard Space Flight Center, Greenbelt, MD 20771, USA\\
$^{19}$Department of Astronomy, University of Maryland, College Park, MD 20742, USA\\
$^{20}$Institute of Natural and Mathematical Sciences, Massey University, Auckland 0745, New Zealand\\
$^{21}$Department of Earth and Planetary Science, Graduate School of Science, The University of Tokyo, 7-3-1 Hongo, Bunkyo-ku, Tokyo 113-0033, Japan\\
$^{22}$Instituto de Astrof\'isica de Canarias, V\'ia L\'actea s/n, E-38205 La Laguna, Tenerife, Spain\\
$^{23}$Oak Ridge Associated Universities, Oak Ridge, TN 37830, USA\\ 
$^{24}$Institute of Space and Astronautical Science, Japan Aerospace Exploration Agency, 3-1-1 Yoshinodai, Chuo, Sagamihara, Kanagawa 252-5210, Japan\\
$^{25}$Sorbonne Universit\'e, CNRS, Institut d'Astrophysique de Paris, IAP, F-75014, Paris, France\\
$^{26}$Department of Physics, University of Auckland, Private Bag 92019, Auckland, New Zealand\\
$^{27}$University of Canterbury Mt.\ John Observatory, P.O. Box 56, Lake Tekapo 8770, New Zealand\\
$^{28}$Villanova University, Department of Astrophysics and Planetary Sciences, 800 Lancaster Ave., Villanova, PA 19085, USA
}

\pubyear{2025}

\begin{document}

\label{firstpage}
\pagerange{\pageref{firstpage}--\pageref{lastpage}}
\maketitle
\clearpage

\begin{abstract}

We present observations and analyses of three high-magnification microlensing events: KMT-2022-BLG-0954, KMT-2024-BLG-0697, and MOA-2024-BLG-018. All three exhibit the ``Planet/Binary'' degeneracy, with planetary solutions corresponding to mass ratios in the range $-3.7 < \log q < -2.2$, while the binary solutions yield $\log q > -2.0$. For KMT-2022-BLG-0954, we identify a previously unrecognized degeneracy among planetary solutions, involving different mass ratios and normalized source radii. In all three cases, single-lens binary-source models are excluded. Bayesian analyses suggest that the planetary solutions correspond to gas giants orbiting M/K dwarfs beyond the snow line, while KMT-2022-BLG-0954 also admits an alternative interpretation as a super-Earth orbiting a late-type M dwarf. The binary solutions imply a diverse set of systems, including M-dwarf pairs and M-dwarf–brown-dwarf binaries. A review of known events subject to the ``Planet/Binary'' degeneracy shows that in most cases the degeneracy cannot be resolved through follow-up high-resolution imaging, particularly in the presence of the newly identified degeneracy.

\end{abstract}




\section{Introduction}\label{intro}

To date, more than 250 planets \citep{NASAExo} have been discovered through the gravitational microlensing technique\footnote{\url{http://exoplanetarchive.ipac.caltech.edu}, as of 2025 August 31} \citep{Shude1991,Andy1992}, most of which are located near or beyond the water snow line \citep{snowline} in their planetary systems. Owing to the degeneracy between lens mass and distance, the host and planetary masses remain undetermined for the majority of these events. Nevertheless, light-curve modeling can yield two key parameters directly linked to planetary properties: the planet-to-host mass ratio, $q$, and the projected separation between the planet and host, expressed in units of the Einstein radius, $s$.

Current statistical studies of microlensing planets have primarily focused on the distribution of planet-to-host mass ratios \citep{mufun,Cassan2012,Suzuki2016,Wise,OB160007}, largely due to the presence of two degeneracies that produce similar values of $q$ but substantially different values of $s$. For example, four of the six planets in the first rigorously defined microlensing planetary sample exhibit the so-called ``close/wide'' degeneracy \citep{mufun}, for which two planetary models are related approximately by $s \leftrightarrow s^{-1}$ \citep{Griest1998, Dominik1999, An2005}. Another common degeneracy is the ``inner/outer'' degeneracy \citep{Gaudi&Gould1997ApJ_continuous_degeneracy}, for which the source passes either inside (``inner'') or outside (``outer'') the planetary caustics relative to the central caustic. In recent years, efforts by \cite{OB190960,KB190253,KMT2021_mass1,Zhang2022} have unified these two types of degeneracies, and \cite{ZhangGaudi2022} has provided a comprehensive theoretical framework to describe them.

Unlike the degeneracies described above, the ``Planet/Binary'' degeneracy produces substantial differences in both $q$ and $s$. This degeneracy, which usually occurs in high-magnification (HM) events\footnote{One exception is the event KMT-2016-BLG-1855 \citep{2016_prime}}, was first identified by \citet{HanGaudi2008}, and \citet{OB110950} presented the first realistic examples. In such cases, a double-peaked anomaly can arise from two cusp approaches to the central caustic, which may be generated either by a planetary system or by a stellar binary. Because HM events are particularly sensitive to planetary perturbations \citep{Griest1998}, this degeneracy can introduce contamination into planetary statistical samples \citep{ShangYang2026}. For example, although the Korea Microlensing Telescope Network (KMTNet; \citealt{KMT2016}) relies less on follow-up observations for HM events than earlier programs (e.g., \citealt{mufun, Suzuki2016}), in the 63-planet sample of KMTNet, three events affected by the ``Planet/Binary'' degeneracy, KMT-2018-BLG-2164, OGLE-2018-BLG-1554, and KMT-2018-BLG-2718 \citep{Paper5Gould2022}, were excluded from the mass-ratio function analysis, reducing the sample by 5\%, given the 63 surviving planets \citep{OB160007}.

Since July 2020, the Microlensing Astronomy Probe (MAP\footnote{\url{http://i.astro.tsinghua.edu.cn/~smao/MAP/}}) collaboration has been using the Las Cumbres Observatory global network (LCOGT, \citealt{LCOGT}) to follow up KMTNet HM events, coordinated with the KMTNet ``auto-followup'' system and the Microlensing Follow Up Network ($\mu$FUN, \citealt{mufun}). This follow-up program aims to establish a statistical sample of microlensing planets. To date, the collected data have contributed to the analysis of 12 planetary events, including five low mass-ratio planets with $q < 10^{-4}$ \citep{KB200414, KB210171, KB210912, KB220440, KB231431}, as well as two lensing systems hosting two planet-like companions \citep{KB200414,KB221818}.

In this paper, we analyze three events from this follow-up program that exhibit the ``Planet/Binary'' degeneracy. The structure of the paper is as follows. In Section~\ref{obser}, we describe the survey and follow-up observations. Sections~\ref{2L1S} and~\ref{1L2S} present the binary-lens single-source (2L1S) and single-lens binary-source (1L2S) analyses, respectively. In Section~\ref{CMD}, we show the color-magnitude diagram (CMD) and derive the source properties. The lens physical properties, estimated through a Bayesian analysis, are presented in Section~\ref{lens}. Finally, in Section~\ref{dis}, we place these three events in the context of other known cases exhibiting the ``Planet/Binary'' degeneracy, and we assess the prospects of resolving this degeneracy through light-curve analysis, satellite parallax, and follow-up high resolution imaging.

\section{Observations and Data Reduction}\label{obser}

\begin{table*}
    \renewcommand\arraystretch{1.5}
    \centering
    \caption{Event Names, First Alert Dates, Locations, and Survey Cadences for the three events}
    \begin{tabular}{c c c c c c c}
    \hline
    \hline
    Event Name & Alert Date & ${\rm RA}_{\rm J2000}$ & ${\rm Decl.}_{\rm J2000}$ & $\ell$ & $b$ & $\Gamma$ \\
    \hline 
    \eventa & 25 May 2022 & 18:00:07.36 & $-$29:02:04.49 & 1.53 & $-$2.78 & $3.0~{\rm hr}^{-1}$  \\
    MOA-2022-BLG-307 & & & & & & $3.0~{\rm hr}^{-1}$  \\
    \hline 
    \eventb & 22 Apr 2024 & 17:43:27.42 & $-$24:53:10.86 & +3.20 & +2.53 & $1.0~{\rm hr}^{-1}$ \\
    \hline
    \eventc & 28 May 2024 & 17:51:23.88 & $-$34:20:02.29 & $-$4.00 & $-$3.83 & $1.3~{\rm hr}^{-1}$  \\
    KMT-2024-BLG-1170 & & & & & & $1.0~{\rm hr}^{-1}$  \\
    OGLE-2024-BLG-0746 & & & & & & $0.5~{\rm day}^{-1}$ \\
    \hline
    \hline
    \end{tabular}
    \label{event_info}
\end{table*}

\subsection{Survey Observations}

Two of the events, \eventa\ and \eventb, were first discovered by the KMTNet alert-finder system \citep{KMTAF}. The former was also independently alerted by the Microlensing Observations in Astrophysics (MOA; \citealt{Sako2008}) group as MOA-2022-BLG-307. The third event, MOA-2024-BLG-018, was initially discovered by MOA and subsequently alerted by KMTNet as KMT-2024-BLG-1170 and the Early Warning System \citep{Udalski1994,Udalski2003} of the Optical Gravitational Lensing Experiment (OGLE; \citealt{OGLEIV}) as OGLE-2024-BLG-0746. Throughout this paper, we refer to the events by their first-discovery names.

KMTNet observations were conducted using three identical 1.6 m telescopes, each equipped with a $4~{\rm deg}^2$ camera, located in Chile (KMTC), South Africa (KMTS), and Australia (KMTA). The field placement and observing cadences for KMTNet are detailed in Figure 12 of \cite{KMTeventfinder}. The MOA group employed the 1.8 m telescope, featuring a $2.2~{\rm deg}^2$ camera, at Mt. John University Observatory in New Zealand. Meanwhile, the OGLE group collected data with the 1.3 m Warsaw Telescope, which is equipped with a $1.4~{\rm deg}^2$ field-of-view camera, at Las Campanas Observatory in Chile. A summary of the event names, alert dates, equatorial and Galactic coordinates, and survey cadences for the three events is provided in Table~\ref{event_info}.

Most KMTNet and OGLE observations were conducted in the $I$ band, while MOA data were primarily obtained in the MOA-Red band, which approximately corresponds to the combined standard Cousins $R$ and $I$ bands. In all three surveys, a subset of $V$-band images was also acquired to enable source color measurements. For \eventa, the V-band data from the BLG43 field, observed by KMTC and KMTS, and from the BLG04 field, observed by KMTS, help exclude the 1L2S model and are therefore included in the light curve analysis. 

\subsection{Follow-up Observations}

For all three events, follow-up observations were initiated in response to alerts from the KMTNet HighMagFinder system \citep{KB210171}, which evaluates KMTNet events every three hours and aims to issue alerts before the magnification exceeds the follow-up threshold of $A_{\rm thresh} = 25$.

For \eventa, HighMagFinder issued the alert at UT 05:07 on 2022-06-05 (${\rm HJD}^{\prime} = 9735.7$, where ${\rm HJD}^{\prime} \equiv {\rm HJD} - 2450000$), approximately 4.7 days before the peak magnification. Follow-up observations were then carried out using the 1.0-meter LCOGT telescopes in Chile (LCOC) and South Africa (LCOS). 

For \eventb, the alert was issued at UT 14:04 on 2024-04-24 (${\rm HJD}^{\prime} = 10425.58$), about 0.2 days before the peak magnification. Follow-up observations were then conducted with the LCOC and LCOS telescopes and a 0.6 m telescope at Observatorio do Pico dos Dias (OPD) in Brazil. In addition, KMTNet increased the cadence at KMTC by switching from BLG41 to BLG19 observations, so the combined KMTC cadence during the peak is $3~{\rm hr}^{-1}$. 

For \eventc, HighMagFinder issued the alert at UT 06:12 on 2024-05-30 (${\rm HJD}^{\prime} = 10460.75$), at the time of peak. Follow-up observations were immediately initiated with the LCOC telescope.

The LCOGT and OPD images were obtained in the $I$ band.

\subsection{Data Reduction}

The data used in the light-curve analysis were reduced using various difference image analysis (DIA) pipelines \citep{Tomaney1996,Alard1998}: pySIS \citep{pysis,pysis5} for the KMTNet, LCOGT and OPD data, the pipeline of \cite{Bond2001} for the MOA data, and that of \cite{Wozniak2000} for the OGLE data. The error bars from each DIA pipeline were rescaled following the method of \cite{MB11293}, which adjusts the error bars to ensure that the $\chi^2$ per degree of freedom (dof) for each data set is unity.

\section{Binary-lens Single-source Analysis}\label{2L1S}

\subsection{Preamble}
\label{subsection_2L1S_Preamble}

The observed light curves of all three events exhibit anomalies near the peak that deviate from the single-lens single-source (1L1S) model \citep{Paczynski1986}. Such anomalies can potentially be explained by 2L1S configurations. Therefore, in this section, we conduct a 2L1S analysis for the three events.

A standard 2L1S model requires seven parameters to describe the time-dependent magnification, $A(t)$. The first four parameters, $t_0$, $u_0$, $\te$, and $\rho$, are the same as those used in the 1L1S model. Specifically, $t_0$ is the time of the source's closest approach to a reference point in the lens system (e.g., the position of the lens in the 1L1S model); $u_0$ is the impact parameter of this approach in units of the angular Einstein radius $\thetae$; $\te$ is the Einstein radius crossing time; and $\rho$ is the angular source radius $\theta_*$ normalized by $\thetae$ ($\rho = \theta_*/\thetae$). The choice of reference point in the lens system adopted in this work will be specified later. The timescale $\te$ is related to the lens mass $M_{\rm L}$ and $\thetae$ by:
\begin{equation}\label{eqn:te} 
\te = \frac{\thetae}{\mu_{\rm rel}}; \qquad \thetae = \sqrt{\kappa \Ml \pirel}; \qquad \kappa \equiv \frac{4G}{c^2\mathrm{au}} \simeq 8.144 \frac{{\rm mas}}{M_{\odot}},
\end{equation}
where $\Ml$ is the total mass of the lens system, $\mu_{\rm rel}$ is the lens-source relative proper motion in the geocentric frame, and $\pirel = {\rm au} (\Dl^{-1}-\Ds^{-1})$ is the lens-source relative parallax, with $\Dl$ and $\Ds$ denoting the lens and source distances, respectively.

The remaining three parameters ($q$, $s$, $\alpha$) define the binary geometry: $q$ is the mass ratio between the secondary and the primary lenses, $s$ is the projected separation between the binary lenses normalized to $\thetae$, and $\alpha$ is the angle between the source trajectory and the binary axis. 
In addition to the seven parameters that determine the magnification, we introduce two flux parameters ($f_{{\rm S},i}$, $f_{{\rm B},i}$) for each data set $i$ to represent the source flux and any blended flux, respectively. These latter accounts for contamination from unrelated stars or potential lens flux. The observed flux $f_{i}(t)$ is then modeled as
\begin{equation}
    f_{i}(t) = f_{{\rm S},i} \times A(t) + f_{{\rm B},i}. 
\end{equation}

We employ the latest version of the advanced contour integration code, \texttt{VBMicrolensing} \citep{Bozza2010,Bozza2018,VBMicrolensing2025}, to compute the 2L1S magnification.

\begin{figure*}
    \includegraphics[width=1.05\textwidth]{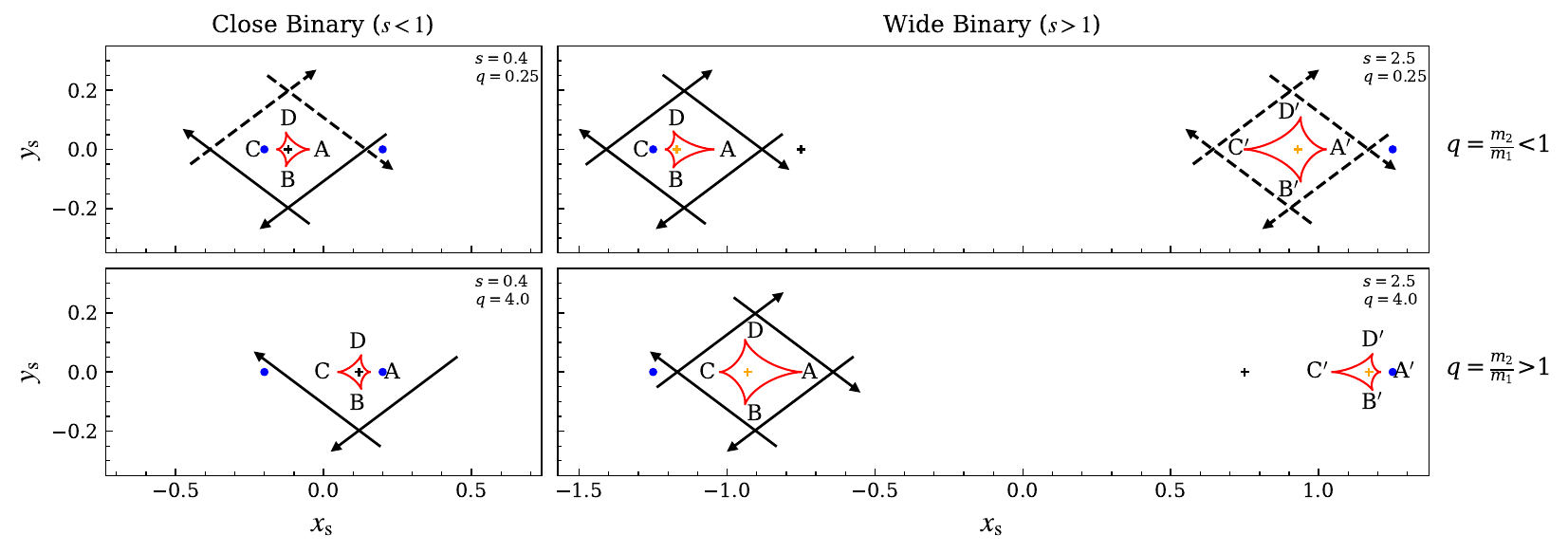}
    \caption{
    Schematic diagram of lens configurations and corresponding caustics for binary models. In each panel, the left and right blue dots represent the primary and secondary lenses with masses $m_1$ and $m_2$, respectively. The left and right columns correspond to models with $s < 1$ and $s > 1$, while the top and bottom rows correspond to $q < 1$ and $q > 1$, where $q \equiv m_2/m_1$. Black and orange crosses mark the center of mass and the magnification centers, respectively. The red lines indicate the caustics, with cusps labeled A–D and $\rm A^{\prime}$–$\rm D^{\prime}$. As described in Section~\ref{subsection_2L1S_Preamble}, because all mass ratios are considered, the solid trajectories are explored in this work, while the dashed trajectories represent equivalent models to those shown with solid trajectories.
    }
\label{Binarycau}
\end{figure*}

All three events exhibit the well-known ``Planet/Binary'' degeneracy. In the planetary interpretation, the anomaly signal arises from the source approaching the two off-axis cusps of the wedge-shaped central caustic. In such cases, the source trajectories are nearly indistinguishable between the close ($s < 1$) and wide ($s > 1$) configurations, leading to the so-called ``Close/Wide'' degeneracy \citep{Griest1998, Dominik1999, An2005}. In addition to the ``Close/Wide'' degeneracy, the geometry of the binary models is more complex. The anomaly signal could result from the source approaching any pair of consecutive cusps on the diamond-shaped caustic, so multiple orientations of the source trajectory are possible. 

Due to this added complexity, we illustrate the source trajectories and caustic structures of the binary models in Figure \ref{Binarycau}, and we define the distinct configurations prior to conducting a detailed numerical analysis. The figure is arranged in a $2\times2$ grid: the left and right columns correspond to the close ($s < 1$) and wide ($s > 1$) models, respectively, while the upper and lower rows correspond to cases with $q < 1$ and $q > 1$, respectively. In all panels, the lens component used as the reference for the mass ratio (that is, the primary lens) is positioned on the left side of the mass center.

For close binary models, we label the four cusps of the central caustic as ``A'', ``B'', ``C'', and ``D'', as shown in Figure \ref{Binarycau}. While there are eight possible orientations for the source trajectory connecting pairs of consecutive cusps, the axial symmetry of the binary lens system reduces these to four non-redundant configurations. We define these orientations as ``AB'', ``BC'', ``CD'', and ``DA'', where the trajectory direction is from the first cusp to the second. These correspond to the solid and dashed trajectories illustrated in the upper-left panel of Figure \ref{Binarycau}. If we restrict the parameter space to $q < 1$, all four orientations must be considered to fully explore the parameter space. However, in this work, we allow $q > 1$ as well. Because the caustic structure is invariant under the transformation $q \leftrightarrow 1/q$ (aside from an inversion of lens positions), the number of required trajectory orientations can be further reduced to two: one selected from ``AB'' or ``CD'', and the other from ``BC'' or ``DA'' \citep{Bozza2016}. 
In this work, ``AB'' and ``BC'' are chosen, which are shown as solid trajectories in the left panels of Figure \ref{Binarycau}. 

Wide binary models present additional complexity due to the presence of two separate diamond-shaped caustics. We retain the notations ``A'', ``B'', ``C'', and ``D'' to label the four cusps of the caustic near the primary lens, and introduce ``$\rm A^{\prime}$'', ``$\rm B^{\prime}$'', ``$\rm C^{\prime}$'' and ``$\rm D^{\prime}$'' for the caustic near the secondary lens. Under the constraint $q < 1$, a total of eight source trajectory orientations around both caustics must be considered (solid and dashed trajectories in the upper-right panel of Figure \ref{Binarycau}). However, because we allow $q > 1$ in this work, the symmetry of the caustic structure under the transformation $q \leftrightarrow 1/q$ permits us to consider only the four orientations around the caustic near the primary lens (solid trajectories in the right panels of Figure \ref{Binarycau}). They are ``AB'', ``BC'', ``CD'', and ``DA''. Combining planetary, close binary, and wide binary models theoretically yields a total of $2 + 2 + 4 = 8$ models for events with the ``Planet/Binary'' degeneracy. 

The centers of the diamond-shaped caustics, defined as the magnification centers, are different for close and wide binary models. In the case of close binary models, the magnification center coincides with the center of mass. In contrast, for wide binary models, the magnification centers are located near each lens component, with shifts induced by the gravitational influence of the other lens. These shifts can be described analytically as \citep{DiStefanoMao1996, Dominik1999, AnHan2002, Chung2005}:
\begin{equation}\label{equ:magnificationcenter}
\begin{array}{ll}
~~~ x_{\rm magnification} ~~ = x_{\rm mass} & (s < 1) \\[2.5ex]
\left.
  \begin{array}{l}
    x_{\rm magnification, 1} = x_{1} + \dfrac{1}{s} \dfrac{q}{1+q} \\[2ex]
    x_{\rm magnification, 2} = x_{2} - \dfrac{1}{s} \dfrac{1}{1+q}
  \end{array}
\right\} & (s > 1). 
\end{array}
\end{equation}
Here, $x_{1}$ and $x_{2}$ denote the coordinates of the primary and secondary lenses, respectively, while $x_{\rm mass}$ refers to the coordinate of the center of mass. In this work, we adopt the magnification center as the reference point for $t_0$ and $u_0$: specifically, we use $x_{\rm magnification}$ for close models, and the magnification center near the primary lens, $x_{\rm magnification, 1}$, for wide models.

To comprehensively explore the 2L1S parameter space and identify all viable local $\chi^2$ minima, we conduct a grid search \citep{Dong2009} over $(\log s, \log q, \log \rho, \alpha)$. Specifically, we sample 61 evenly spaced values over $-1.5 \leq \log s \leq 1.5$, nine evenly spaced values over $-4.0 \leq \log \rho \leq -1.6$, and 16 evenly spaced initial values over $0 \leq \alpha < 2\pi~\mathrm{radians}$. For $\log s < 0$, because the parameter space with $\log q < 0$ and all $\alpha$ adequately covers all solutions, we adopt 61 evenly spaced values in $-6 \leq \log q \leq 0$. For $\log s > 0$, we adopt 101 evenly spaced values in $-6 \leq \log q \leq 4$. The initial values of $(t_0, u_0, \te)$ are taken from the best-fit 1L1S model of each event. At each grid point, a $\chi^2$ minimization is performed using the \texttt{emcee} ensemble sampler \citep{emcee}, with $(\log s, \log q, \log \rho)$ held fixed, while $(t_0, u_0, \te, \alpha)$ are allowed to vary.

After the grid search, one or more local minima in the $(\log s, \log q, \log \rho, \alpha)$ space might be revealed. We then refine them by MCMC with all seven parameters of the standard 2L1S model free to explore the posterior distributions, and thus estimate the uncertainties of the parameters. The final step involves a $\chi^2$ minimization using the Nelder–Mead simplex algorithm\footnote{Implemented via \texttt{scipy.optimize.fmin}. See \url{https://docs.scipy.org/doc/scipy/reference/generated/scipy.optimize.fmin.html\#scipy.optimize.fmin}} to further improve the fit around each minimum. 

We also investigate whether the microlensing parallax vector \citep{Gould1992, Gould2000, Gouldpies2004}
\begin{equation}\label{equ:pie}
    \bm{\pi}_{\rm E} \equiv \frac{\pi_{\rm rel}}{\thetae} \frac{\bm{\mu}_{\rm rel}}{\mu_{\rm rel}},
\end{equation}
can be meaningfully constrained by the data. This vector is parameterized by its two components, $\pi_{\rm E,N}$ and $\pi_{\rm E,E}$, which correspond to the north and east components in equatorial coordinates. In addition to microlensing parallax, we consider the effects of lens orbital motion \citep{MB09387, OB09020} when modeling events with parallax. Furthermore, we fit both the $u_0 > 0$ and $u_0 < 0$ solutions to account for the well-known ecliptic degeneracy \citep{Jiang2004, Poindexter2005}. 

The detailed 2L1S analyses for the three events are presented
separately in the following. 

\subsection{KMT-2022-BLG-0954}
\label{subsection_kb220954}

\begin{figure}
    \centering
    \includegraphics[width=\columnwidth]{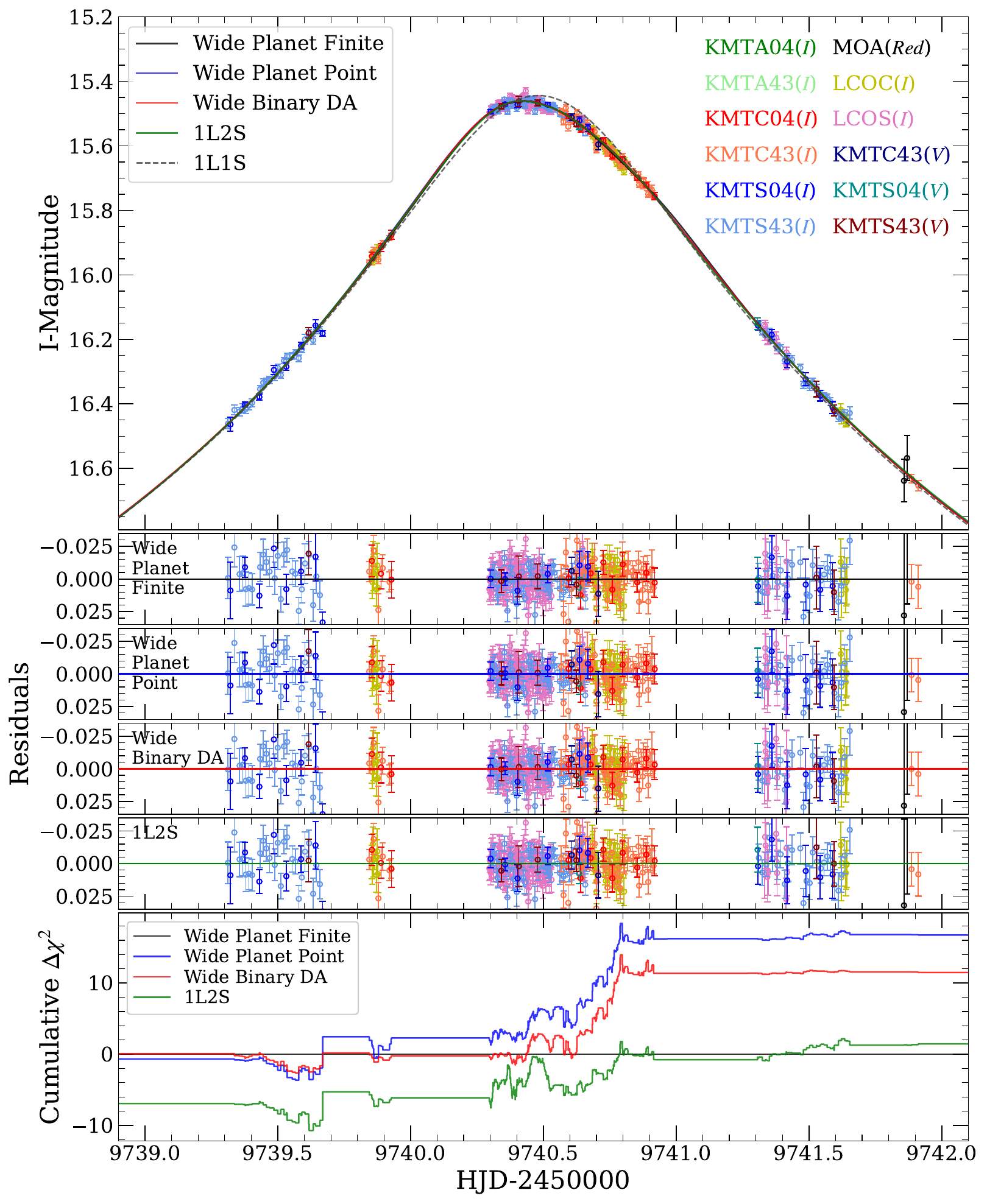}
    \caption{Observed light curves and lensing models for \eventa. The asymmetric peak deviates from the 1L1S model (dashed line). Different data sets are plotted in different colors. The 2L1S model curves correspond to the best ``Planet Finite'', ``Planet Point'', and ``Binary'' solutions. Residuals with respect to the models are also shown. The bottom panel presents the cumulative $\Delta\chi^2$ distribution of the solutions relative to the best-fit 2L1S solution, that is, the ``Wide Planet Finite'' solution. The lensing parameters for all 2L1S solutions are given in Table~\ref{KB220954_parm_2L1Sstatic}.}
    \label{KB220954_lc}
\end{figure}

\begin{figure}
    \centering
    \includegraphics[width=\columnwidth]{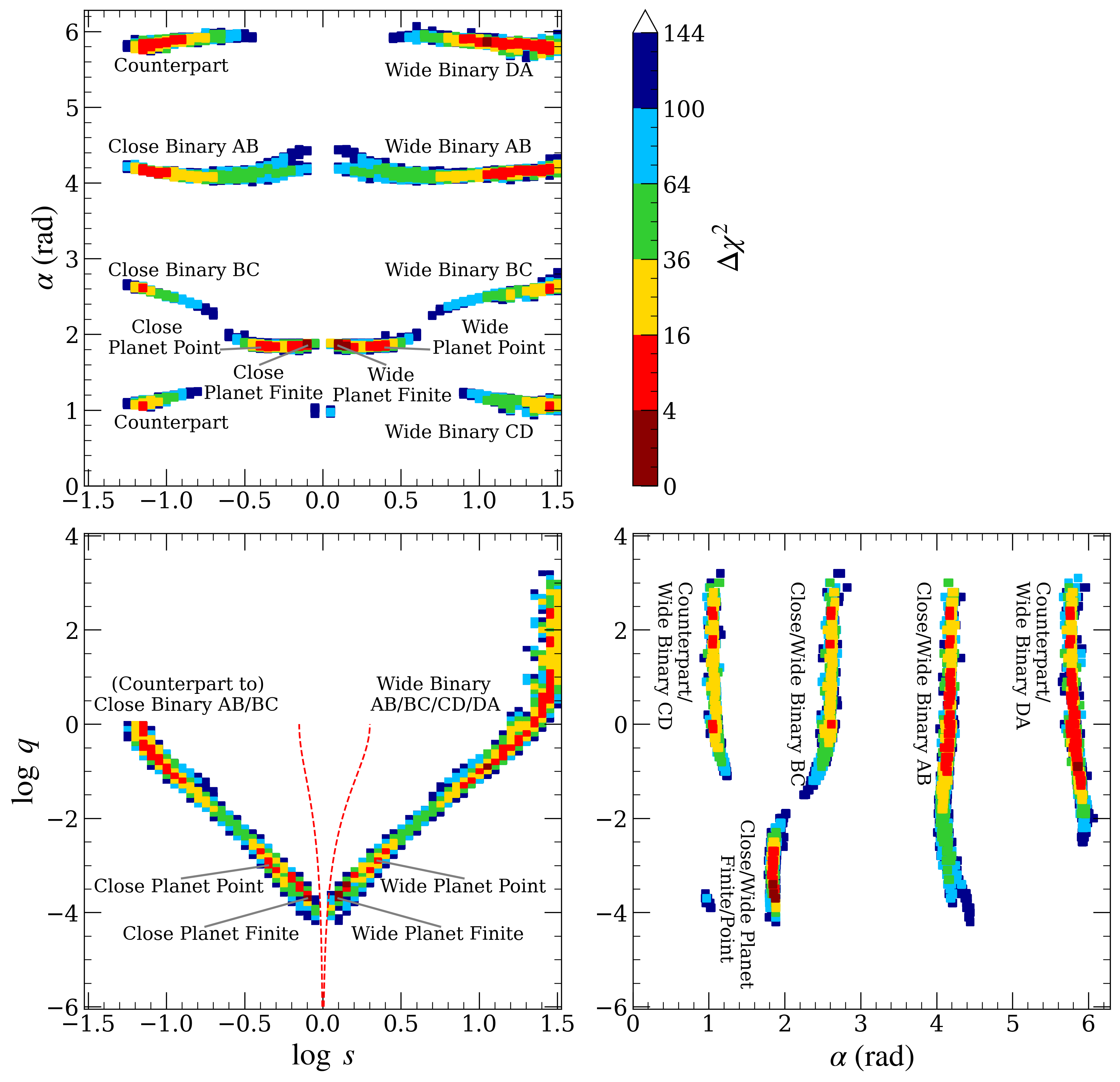}
    \caption{$\chi^2$ surface in the $(\log s, \log q, \alpha)$ space from the grid search of \eventa. Dark red, red, yellow, green, blue, and dark blue indicate grid points within $<1n\sigma$, $<2n\sigma$, $<3n\sigma$, $<4n\sigma$, $<5n\sigma$, and $<6n\sigma$, respectively, where $n^2 = 4$. Grid points with $>6n\sigma$ are left blank. The 12 identified local minima are labeled with their corresponding names. The two red dashed lines mark the boundaries between resonant and non-resonant caustics, following Equation (60) and (61) of \citet{Dominik1999}.}
    \label{KB220954_gridsearch}
\end{figure}

\begin{figure}
    \includegraphics[width=0.47\textwidth]{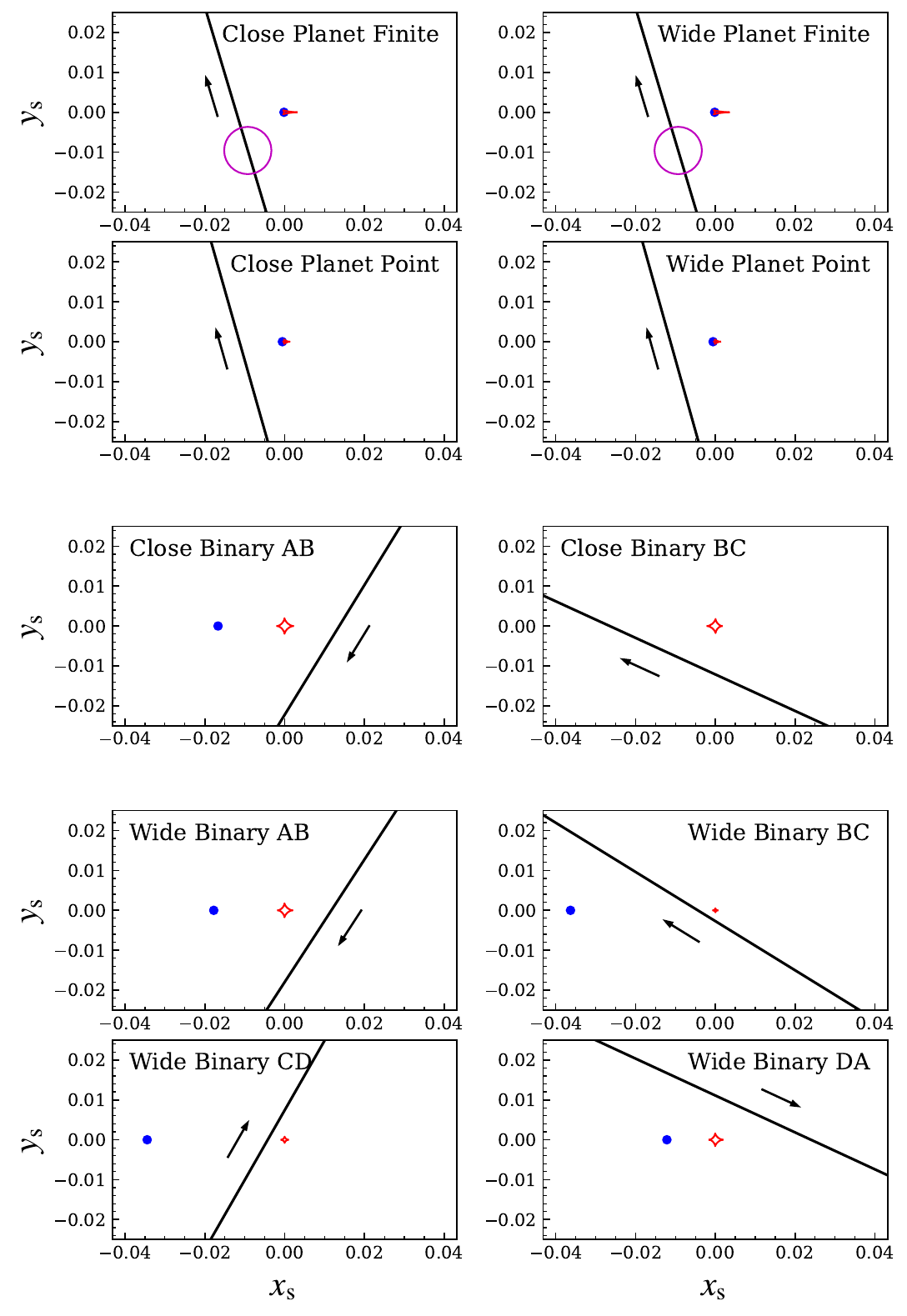}
    \caption{Caustic geometries of the ten 2L1S models for \eventa. The models are grouped into three categories, separated by spacing: four ``Planet'' models, two ``Close Binary'' models, and four ``Wide Binary'' models. In each panel, the red curves denote the caustic, the blue dot marks the position of the primary lens, the coordinate origin is defined at the magnification center, and the black line shows the source–lens relative trajectory, with the arrow indicating the direction of source motion. For the ``Close Planet Finite'' and ``Wide Planet Finite'' models, $\rho$ is well constrained, and the magenta circles represent the angular source radius.}
\label{KB220954_caustic}
\end{figure}

\begin{table*}
    \renewcommand\arraystretch{1.25}
    \centering
    \caption{2L1S Parameters for KMT-2022-BLG-0954}
    
    
    \begin{tabular}{c|c|c|r r r r r r r r}
    \hline
    \hline
    \multicolumn{2}{c|}{Model} & 
    $\chi^2$/dof & 
    $t_{0}$ (${\rm HJD}^{\prime}$) &
    $u_{0} $ &
    $\te$ (days) &
    $\rho (10^{-3})$ &
    $\alpha$ (rad) &
    $s$ &
    $\log q$ &
    $I_{\rm S, OGLE}$ \\
    
    \hline
    \multirow{8}{*}{Close} 
    
    & {Planet Finite}
    & $9785.4/9797$
    & $9740.4804$ & $0.0116$ & $44.73$ & $6.32$ & $1.854$ & $0.764$ & $-3.63$ & $20.33$ \\
    & & & $0.0011$ & $0.0002$ & $0.86$ & $0.27$ & $0.014$ & $0.042$ & $0.11$ & $0.02$ \\     

    \cline{2-11}
    & {Planet Point}
    & ${9802.1/9797}$ 
    & ${9740.4815}$ & ${0.0109}$ & ${45.49}$ & ${<5.01}$ & ${1.845}$ & ${0.438}$ & ${-2.90}$ & ${20.35}$ \\
    & & & ${0.0013}$ & ${0.0002}$ & ${0.88}$ & ${-}$ & ${0.013}$ & ${0.025}$ & ${0.10}$ & ${0.02}$ \\

    \cline{2-11}
    & Binary AB 
    & $9803.9/9797$
    & $9740.4793$ & $0.0115$ & $43.91$ & $<5.25$ & $4.155$ & $0.080$ & $-0.64$ & $20.31$ \\
    & & & $0.0014$ & $0.0002$ & $0.80$ & $-$ & $0.023$ & $0.012$ & $0.23$ & $0.02$ \\

    \cline{2-11}
    & {Binary BC}
    & ${9797.8/9797}$
    
    & ${9740.4802}$ & ${0.0111}$ & ${45.11}$ & ${<6.31}$ & ${2.706}$ & ${0.089}$ & ${0.81}$ & ${20.34}$ \\
    
    & & & ${0.0012}$ & ${0.0002}$ & ${0.91}$ & ${-}$ & ${0.023}$ & ${0.012}$ & ${0.19}$ & ${0.02}$ \\
    \hline
    \multirow{12}{*}{Wide} 
    & \textbf{Planet Finite}
    & $\boldsymbol{9784.5/9797}$

    & $\boldsymbol{9740.4806}$ & $\boldsymbol{0.0116}$ & $\boldsymbol{44.76}$ & $\boldsymbol{6.30}$ & $\boldsymbol{1.856}$ & $\boldsymbol{1.297}$ & $\boldsymbol{-3.64}$ & $\boldsymbol{20.33}$ \\
    
    & & & $\boldsymbol{0.0012}$ & $\boldsymbol{0.0002}$ & $\boldsymbol{0.86}$ & $\boldsymbol{0.27}$ & $\boldsymbol{0.013}$ & $\boldsymbol{0.080}$ & $\boldsymbol{0.11}$ & $\boldsymbol{0.02}$ \\

    \cline{2-11}
    & {Planet Point}
    & ${9802.1/9797}$
    
    & ${9740.4816}$ & ${0.0109}$ & ${45.63}$ & ${<5.25}$ & ${1.847}$ & ${2.295}$ & ${-2.88}$ & ${20.35}$ \\
    
    & & & ${0.0012}$ & ${0.0002}$ & ${0.87}$ & ${-}$ & ${0.014}$ & ${0.119}$ & ${0.10}$ & ${0.02}$ \\

    \cline{2-11}
    & Binary AB 
    & $9803.1/9797$
    
    & $9740.4821$ & $0.0101$ & $50.46$ & $<5.89$ & $4.141$ & $15.587$ & $-0.55$ & $20.31$ \\

    & & & $0.0013$ & $0.0005$ & $2.75$ & $-$ & $0.018$ & $2.694$ & $0.23$ & $0.02$ \\

    \cline{2-11}
    & Binary BC 
    & $9807.4/9797$
    
    & $9740.4755$ & $0.0023$ & $225.51$ & $<1.91$ & $2.598$ & $27.170$ & $1.41$ & $20.30$ \\
    
    & & & $0.0011$ & $0.0002$ & $18.79$ & $-$ & $0.016$ & $1.050$ & $0.08$ & $0.02$ \\

    \cline{2-11}
    & Binary CD 
    & $9808.8/9797$
    
    & $9740.4781$ & $0.0039$ & $134.70$ & $<2.88$ & $1.062$ & $27.210$ & $0.92$ & $20.30$ \\
    
    & & & $0.0011$ & $0.0003$ & $11.44$ & $-$ & $0.017$ & $1.289$ & $0.09$ & $0.02$ \\

    \cline{2-11}
    & {Binary DA}
    & ${9797.8/9797}$

    & ${9740.4783}$ & ${0.0100}$ & ${49.90}$ & ${<5.25}$ & ${5.853}$ & ${13.954}$ & ${-0.69}$ & ${20.34}$ \\
    
    & & & ${0.0012}$ & ${0.0004}$ & ${1.96}$ & ${-}$ & ${0.019}$ & ${2.397}$ & ${0.20}$ & ${0.02}$ \\

    \hline
    \hline
    \end{tabular}
    
    \begin{tablenotes}
        \centering
        \item{NOTE. } 
        ${\rm HJD}^{\prime} = {\rm HJD} - 2450000$. The parameters are reported with their $1\sigma$ uncertainties, while the upper limit on $\rho$ corresponds to $3\sigma$. The values of $t_0$ and $u_0$ are referenced to the magnification center, as defined in Equation~(\ref{equ:magnificationcenter})  and illustrated in Section \ref{subsection_2L1S_Preamble}. The source magnitude is calibrated to the standard $I$-band system using the OGLE-III star catalog. The best-fit solution is highlighted in bold.
    \end{tablenotes}
    \label{KB220954_parm_2L1Sstatic}
\end{table*}

Figure \ref{KB220954_lc} displays the observed data around the peak region of \eventa. The $I$-band magnitude has been calibrated to the standard $I$-band magnitude using the OGLE-III star catalog \citep{OGLEIII}. The peak shows asymmetry, covered by the KMTC, KMTS, LCOC, and LCOS data. 

We first conduct a grid search over $(\log s, \log q, \log \rho, \alpha)$ as described in Section \ref{subsection_2L1S_Preamble}. The initial values of ($t_0$, $u_0$, $\te$) are seeded at the 1L1S fitting values of KMT-2022-BLG-0954, where $t_0 (\hjdPrime) = 9740.48$, $u_0 = 0.010$, $\te = 48$ days. Figure \ref{KB220954_gridsearch} shows the $\chi^2$ surface in the $(\log s, \log q, \alpha)$ space from the grid search. We identify 12 distinct local minima. 

Of these, four lie within the range $-4 < \log q < -2$ and $-0.5 < \log s < 0.5$. Table~\ref{KB220954_parm_2L1Sstatic} lists the corresponding 2L1S parameters derived from the MCMC analysis, and the caustics are shown in Figure~\ref{KB220954_caustic}. These four solutions follow trajectories similar to those of the “Planet” models described in Section \ref{subsection_2L1S_Preamble}. They can be grouped into two pairs of close/wide degeneracies. In the first pair, with $\log q \sim -3.6$, the normalized source radius is well constrained to $\rho = (6.31 \pm 0.27) \times 10^{-3}$. We designate these solutions as ``Close Planet Finite'' and ``Wide Planet Finite''. In the second pair, with $\log q \sim -2.9$, the observed data are consistent with a point-source model within the $1\sigma$ level. We refer to these as ``Close Planet Point'' and ``Wide Planet Point''.
Notably, the degeneracy between the ``Planet Finite'' and ``Planet Point'' models represents a previously unrecognized form of degeneracy. We will discuss this in detail in Section~\ref{dis}.

For the other eight local minima, their source trajectories are similar to the ``Binary'' trajectories illustrated in Figure~\ref{Binarycau}. Among these, four correspond to ``Close'' configurations. According to the definition in Section~\ref{subsection_2L1S_Preamble}, we retain only the ``AB'' and ``BC'' trajectories, as the two minima labeled ``Counterpart'' in Figure~\ref{KB220954_gridsearch} are intrinsically identical to them. We therefore label the two remaining solutions as ``Close AB'' and ``Close BC''. On the other hand, the four ``Wide'' local minima are non-redundant, and we label them as ``Wide Binary AB'', ``Wide Binary BC'', ``Wide Binary CD'', and ``Wide Binary DA''.

Table~\ref{KB220954_parm_2L1Sstatic} presents the 2L1S parameters for the four ``Planet'' and six ``Binary'' solutions. The ``Wide Planet Finite'' model provides the best fit to the data, while the other nine solutions are disfavored by $0.9 < \Delta\chi^2 < 24.3$. Thus, none of the solutions can be decisively ruled out at the $>5\sigma$ level from the light-curve analysis. We will further discuss the relative preference among these solutions in Section~\ref{lens}, in combination with results from a Bayesian analysis based on a Galactic model.  

For all of the ``Binary'' solutions, the finite-source effect is only marginally detected, and the observed data are consistent with a point-source model at the $1\sigma$ level. It is also notable that the four ``Wide Binary'' solutions yield longer time scales than the other solutions, particularly the ``Wide Binary BC'' ($\te = 225.5$) and ``Wide Binary CD'' ($\te = 134.7$) solutions, both of which have $\log q > 0$. This arises because the observed light curve is sensitive not directly to both widely separated bodies, but rather to the body located near the source trajectory. The $\te$ values obtained from the 2L1S fits refer to the timescales of the two-body system. For the body near the source trajectory, the effective timescale is given by $\te/\sqrt{1+q}$, which yields $\sim 44$ days for the four ``Wide Binary'' solutions, consistent with those of the ``Planet'' and ``Close Binary'' solutions.  

We further examine whether the fits can be improved by incorporating higher-order effects. We find that the $\chi^2$ improvement is only $\sim 1$. However, we obtain a useful constraint on the parallax vector: $\pi_{\rm E, \parallel} = 0.00 \pm 0.38$, where $\pi_{\rm E, \parallel} \sim \pi_{\rm E, E}$ is the minor axis of the elliptical parallax contour, approximately aligned with the direction of Earth’s acceleration. In contrast, for the major axis of the parallax contour, $\pi_{\rm E, \bot} \sim \pi_{\rm E, N}$, there is no useful constraint because $\sigma(\pi_{\rm E, \bot}) \sim 1$. We will include $\pi_{\rm E, \parallel}$ in the Bayesian analysis of Section~\ref{lens} to estimate the physical parameters of the lens.


\subsection{KMT-2024-BLG-0697}
\label{subsection_kb240697}

\begin{figure}
    \centering
    \includegraphics[width=\columnwidth]{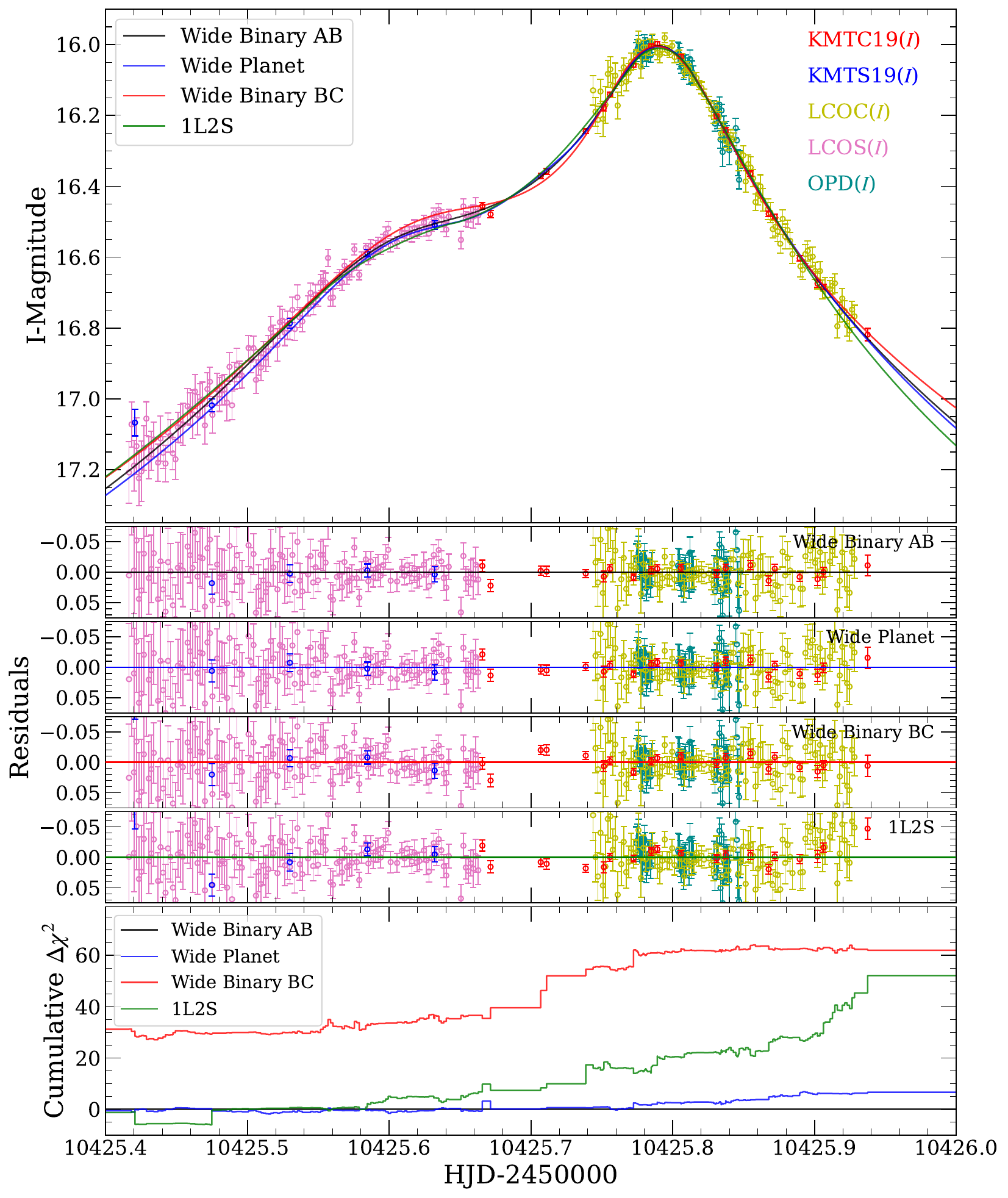}
    \caption{Observed light curves and lensing models for \eventb. Symbols are similar to those in Figure \ref{KB220954_lc}. The ``Wide Binary BC'' model is rejected by $\Delta\chi^2= 66.9$, and the 1L2S model is rejected by $\Delta\chi^2= 57.5$.}
    \label{KB240697_lc}
\end{figure}

\begin{figure}
    \centering
    \includegraphics[width=0.47\textwidth]{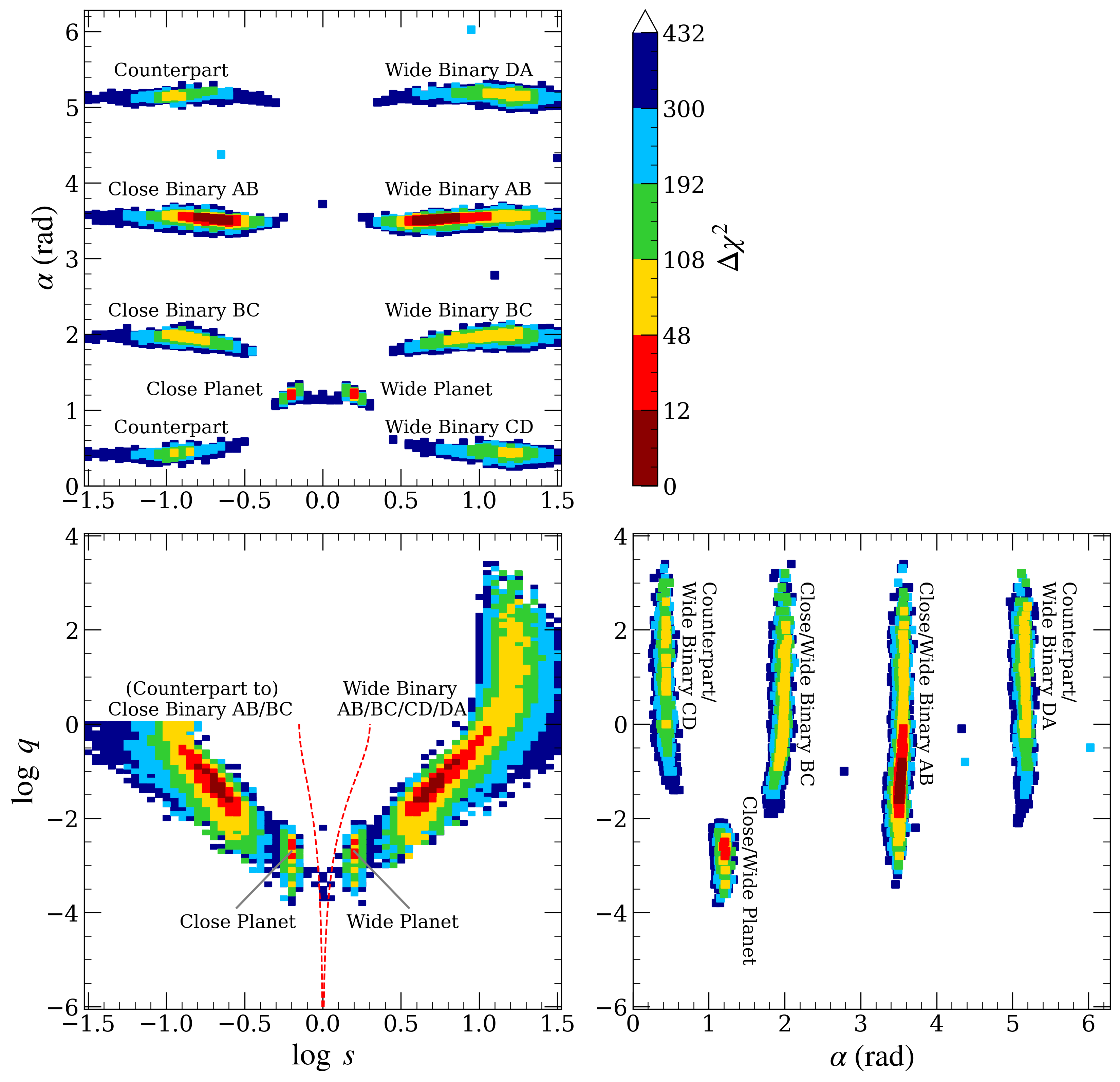}
    \caption{$\chi^2$ surface in the $(\log s, \log q, \alpha)$ space from the grid search of \eventb. The color scheme is the same as in Figure~\ref{KB220954_gridsearch}, except that here $n^2 = 12$. The ten identified local minima are labeled.
    }
\label{KB240697_gridsearch}
\end{figure}

\begin{table*}
    \renewcommand\arraystretch{1.25}
    \centering
    \caption{2L1S Parameters for KMT-2024-BLG-0697}
    
    
    \begin{tabular}{c|c|c|r r r r r r r r}
    \hline
    \hline
    \multicolumn{2}{c|}{Model} & 
    $\chi^2$/dof & 
    $t_{0}$ (${\rm HJD}^{\prime}$) &
    $u_{0} $ &
    $\te$ (days) &
    $\rho (10^{-3})$ &
    $\alpha$ (rad) &
    $s$ &
    $\log q$ &
    $I_{\rm S, KMT}$ \\
     
    \hline
    \multirow{6}{*}{Close} 
    
    & {Planet}
    & ${1086.0/1078}$
    
    & ${10425.7413}$ & ${0.0073}$ & ${14.99}$ & ${2.49}$ & ${1.212}$ & ${0.631}$ & ${-2.62}$ & ${21.48}$ \\
    
    & & & ${0.0009}$ & ${0.0006}$ & ${1.20}$ & ${0.21}$ & ${0.009}$ & ${0.007}$ & ${0.04}$ & ${0.09}$ \\

    \cline{2-11}
    & {Binary AB}
    & ${1078.5/1078}$
    
    & ${10425.7420}$ & ${0.0078}$ & ${14.41}$ & ${2.22}$ & ${3.526}$ & ${0.210}$ & ${-1.31}$ & ${21.42}$ \\
    
    & & & ${0.0008}$ & ${0.0007}$ & ${1.20}$ & ${0.27}$ & ${0.010}$ & ${0.014}$ & ${0.10}$ & ${0.10}$ \\

    \cline{2-11}
    & Binary BC 
    & $1146.0/1078$
    
    & $10425.7486$ & $0.0132$ & $9.97$ & $4.28$ & $1.977$ & $0.134$ & $-0.36$ & $20.90$ \\
    
    & & & $0.0019$ & $0.0008$ & $0.61$ & $0.35$ & $0.015$ & $0.009$ & $0.11$ & $0.07$ \\

    \hline
    \multirow{10}{*}{Wide} 
    & {Planet}
    & ${1085.9/1078}$
    
    & ${10425.7411}$ & ${0.0072}$ & ${15.17}$ & ${2.49}$ & ${1.212}$ & ${1.576}$ & ${-2.62}$ & ${21.49}$ \\
    
    & & & ${0.0009}$ & ${0.0006}$ & ${1.23}$ & ${0.22}$ & ${0.009}$ & ${0.017}$ & ${0.04}$ & ${0.09}$ \\

    \cline{2-11}
    & \textbf{Binary AB}
    & $\boldsymbol{1078.0/1078}$
    
    & $\boldsymbol{10425.7434}$ & $\boldsymbol{0.0076}$ & $\boldsymbol{14.79}$ & $\boldsymbol{2.17}$ & $\boldsymbol{3.522}$ & $\boldsymbol{5.161}$ & $\boldsymbol{-1.27}$ & $\boldsymbol{21.42}$ \\
    
    & & & $\boldsymbol{0.0008}$ & $\boldsymbol{0.0007}$ & $\boldsymbol{1.23}$ & $\boldsymbol{0.28}$ & $\boldsymbol{0.010}$ & $\boldsymbol{0.412}$ & $\boldsymbol{0.11}$ & $\boldsymbol{0.10}$ \\

    \cline{2-11}
    & Binary BC 
    & $1144.9/1078$
    
    & $10425.7470$ & $0.0091$ & $14.61$ & $2.89$ & $1.982$ & $11.663$ & $0.02$ & $20.91$ \\
    
    & & & $0.0015$ & $0.0009$ & $1.52$ & $0.33$ & $0.012$ & $1.096$ & $0.16$ & $0.06$ \\

    \cline{2-11}
    & Binary CD 
    & $1165.0/1078$
    
    & $10425.7444$ & $0.0027$ & $50.81$ & $0.89$ & $0.438$ & $15.884$ & $1.43$ & $20.83$ \\
    
    & & & $0.0014$ & $0.0005$ & $8.40$ & $0.16$ & $0.011$ & $0.462$ & $0.16$ & $0.07$ \\

    \cline{2-11}
    & Binary DA 
    & $1155.7/1078$
    
    & $10425.7410$ & $0.0034$ & $40.73$ & $1.07$ & $5.167$ & $16.094$ & $1.21$ & $20.86$ \\
    
    & & & $0.0014$ & $0.0005$ & $6.43$ & $0.18$ & $0.010$ & $0.452$ & $0.15$ & $0.07$ \\

    
    
    
    \hline
    \hline
    \end{tabular}
    \label{KB240697_parm_2L1Sstatic}
\end{table*}

\begin{figure}
    \includegraphics[width=0.47\textwidth]{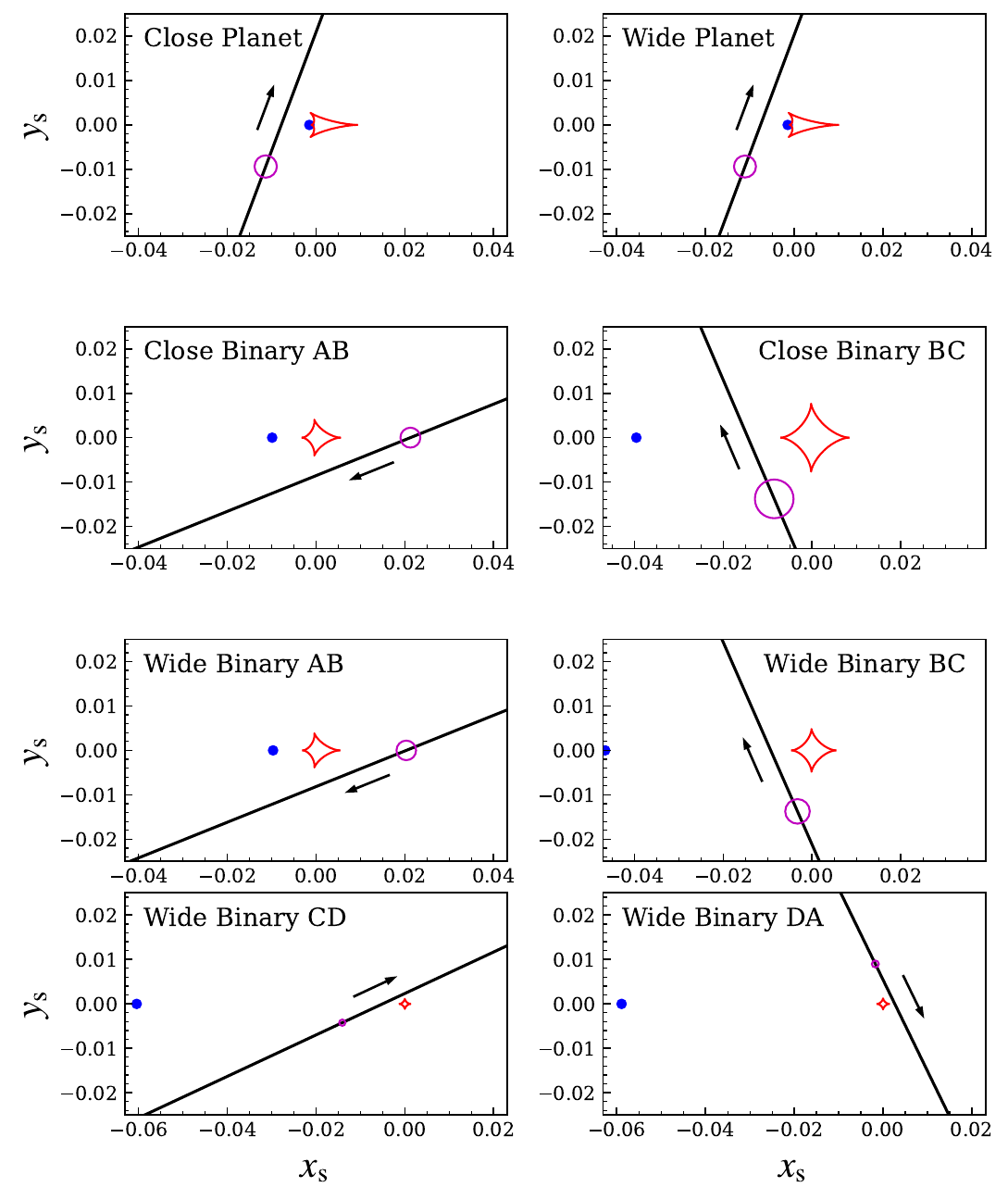}
    \caption{
    Caustic geometries of the eight 2L1S models for \eventb. The symbols follow the same convention as in Figure~\ref{KB220954_caustic}. 
    }
\label{KB240697_caustic}
\end{figure}

Figure~\ref{KB240697_lc} displays the observed data around the peak region of \eventb. The $I$-band magnitude is given in the instrumental system of KMTC, since this event lies outside the OGLE-III footprint. Two distinct bumps are evident in the light curve: the first is covered by KMTS and LCOS data, while the second is covered by KMTC, LCOC, and OPD data.

For the grid search, the initial values of $(t_0, u_0, \te)$ are seeded from the 1L1S fit, where $t_0 (\hjdPrime) = 10425.74$, $u_0 = 0.008$, and $\te = 15$ days. Figure~\ref{KB240697_gridsearch} shows the $\chi^2$ surface in the $(\log s, \log q, \alpha)$ space from the grid search. We identify ten distinct local minima. Similar to \eventa, there are eight ``Binary'' local minima, including four ``Wide Binary'' solutions, two ``Close Binary'' solutions (``Close Binary AB'' and ``Close Binary BC''), and two ``Counterpart'' solutions for ``Close Binary''. For the ``Planet'' configurations, only two solutions are found, which we label ``Close Planet'' and ``Wide Planet''.   

We then refine the eight distinct solutions using both MCMC and the Nelder–Mead simplex algorithm, with all seven parameters of the static 2L1S model free. Table~\ref{KB240697_parm_2L1Sstatic} lists the resulting parameters along with the instrumental $I$-band source magnitude for the eight models, and Figure~\ref{KB240697_caustic} shows their corresponding caustic geometries. Unlike \eventa, finite-source effects are measured in all solutions. Among them, the ``Wide Binary AB'' solution provides the best fit to the light curve. The ``Close Planet,'' ``Close Binary AB,'' and ``Wide Planet'' solutions are disfavored by $\Delta\chi^2 = 8.0, 0.5,$ and $7.9$, respectively, but cannot be definitively ruled out. The other four solutions are disfavored by $\Delta\chi^2 \geq 66.9$ and provide significantly worse fits to the data at both the wings and the peaks of the light curve. Figure~\ref{KB240697_lc} shows an example of the residuals for the ``Wide Binary BC'' solution. Hence, we exclude these four and retain the ``Wide Binary AB'', ``Close Binary AB'', ``Wide Planet'' and ``Close Planet'' solutions for further investigation.

Given the short timescale of this event, the parallax parameters are not expected to be well constrained. Indeed, incorporating higher-order effects improves the fit by only $\Delta\chi^2 = 0.2$, and the $1\sigma$ uncertainty of the parallax vector remains $>1.3$ in all directions. Thus, the constraint on $\pi_{\rm E}$ is not useful for the Bayesian analysis.

\subsection{\eventc}
\label{subsection_kb241170}

\begin{figure}
    \centering
    \includegraphics[width=\columnwidth]{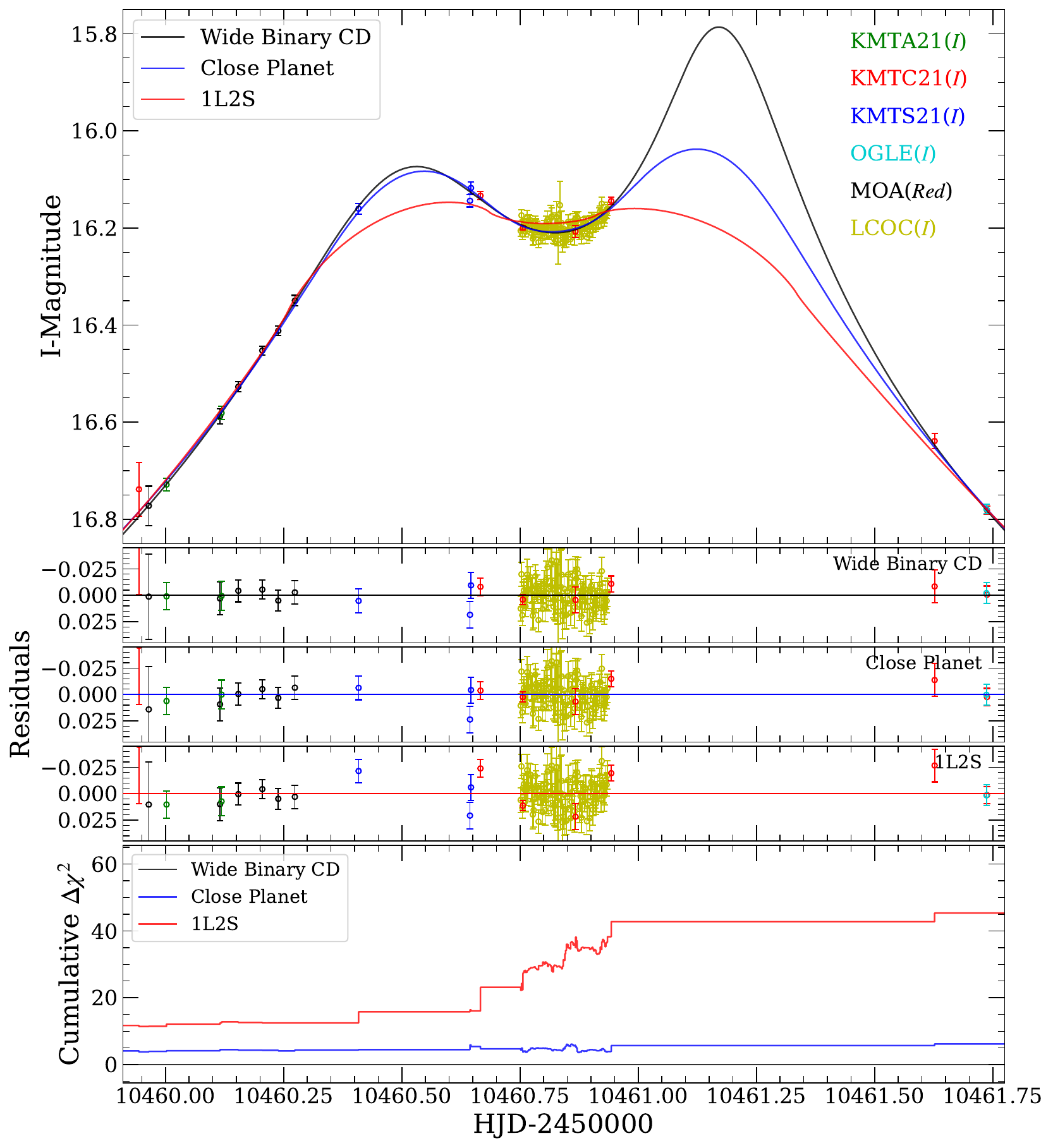}
    \caption{Observed light curves and lensing models for \eventc. Symbols are similar to those in Figure \ref{KB220954_lc}. The 1L2S model is rejected by $\Delta\chi^2= 59.4$.}
    \label{KB241170_lc}
\end{figure}

\begin{figure}
    \centering
    \includegraphics[width=0.47\textwidth]{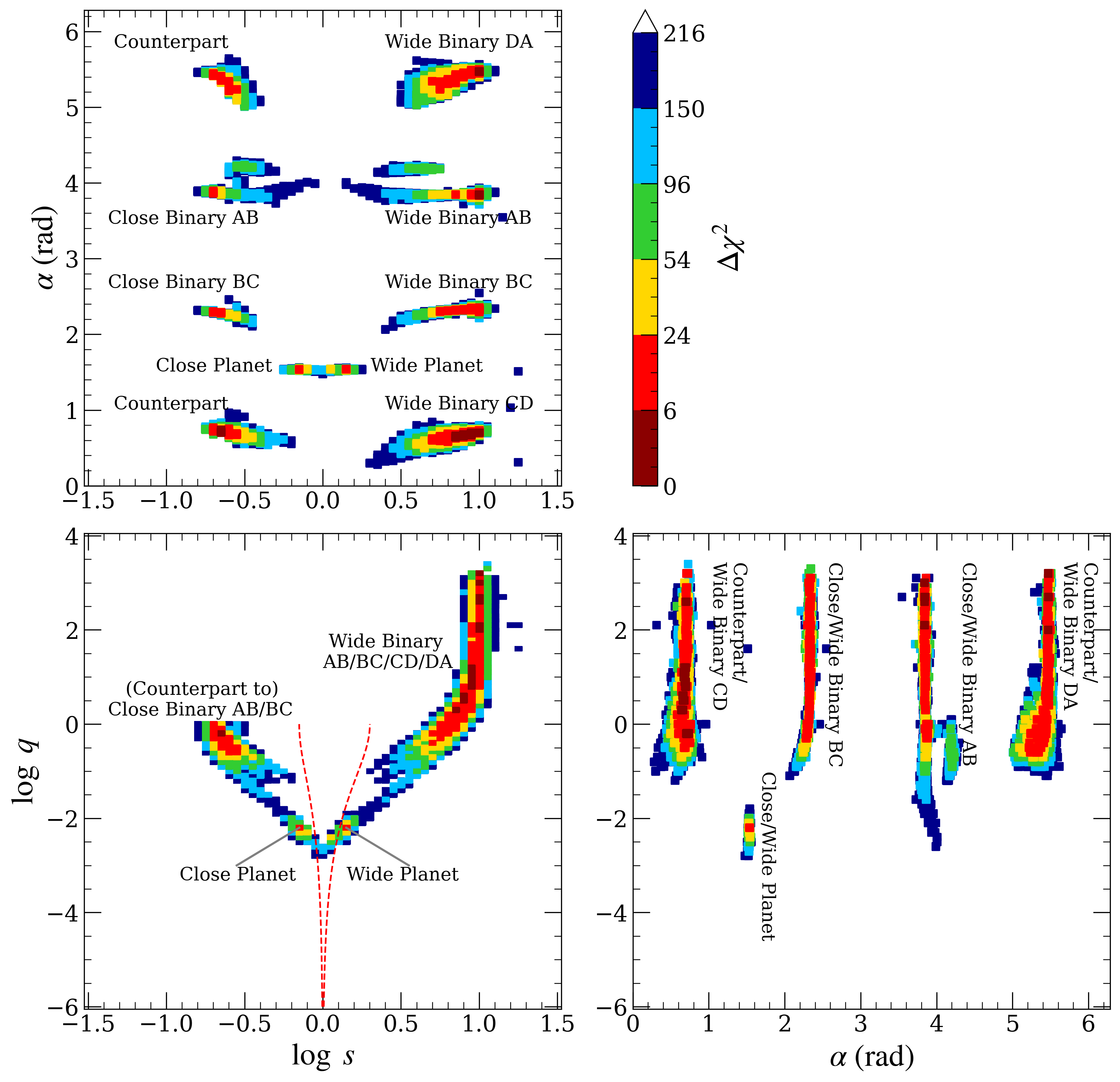}
    \caption{$\chi^2$ surface in the $(\log s, \log q, \alpha)$ space from the grid search of \eventb. The color scheme is the same as in Figure~\ref{KB220954_gridsearch}, except that here $n^2 = 6$. The ten identified local minima are labeled.
    }
\label{KB241170_gridsearch}
\end{figure}

\begin{table*}
    \renewcommand\arraystretch{1.25}
    \centering
    \caption{2L1S Parameters for \eventc}
    
    
    \begin{tabular}{c|c|c|r r r r r r r r}
    \hline
    \hline
    \multicolumn{2}{c|}{Model} & 
    $\chi^2$/dof & 
    $t_{0}$ (${\rm HJD}^{\prime}$) &
    $u_{0} $ &
    $\te$ (days) &
    $\rho (10^{-3})$ &
    $\alpha$ (rad) &
    $s$ &
    $\log q$ &
    $I_{\rm S, KMT}$ \\

    \hline
    \multirow{6}{*}{Close} 
    
    & {Planet}
    & ${1668.7/1651}$
    
    & ${10460.8469}$ & ${0.0317}$ & ${13.92}$ & ${<16.71}$ & ${1.533}$ & ${0.692}$ & ${-2.23}$ & ${19.70}$ \\
    
    & & & ${0.0045}$ & ${0.0013}$ & ${0.48}$ & ${-}$ & ${0.008}$ & ${0.009}$ & ${0.04}$ & ${0.04}$ \\

    \cline{2-11}
    & {Binary AB}
    & ${1651.4/1651}$
    
    & ${10460.8631}$ & ${0.0370}$ & ${12.56}$ & ${<14.13}$ & ${3.854}$ & ${0.226}$ & ${0.22}$ & ${19.52}$ \\
    
    & & & ${0.0091}$ & ${0.0014}$ & ${0.43}$ & ${-}$ & ${0.018}$ & ${0.014}$ & ${0.09}$ & ${0.05}$ \\

    \cline{2-11}
    & Binary BC 
    & $1654.8/1651$
    
    & $10460.8736$ & $0.0364$ & $13.02$ & $<19.50$ & $2.291$ & $0.204$ & $-0.02$ & $19.58$ \\
    
    & & & $0.0065$ & $0.0023$ & $0.49$ & $-$ & $0.010$ & $0.006$ & $0.13$ & $0.05$ \\

    \hline
    \multirow{10}{*}{Wide} 
    & {Planet}
    & ${1669.1/1651}$
    
    & ${10460.8468}$ & ${0.0316}$ & ${13.95}$ & ${<16.90}$ & ${1.533}$ & ${1.409}$ & ${-2.22}$ & ${19.70}$ \\
    
    & & & ${0.0043}$ & ${0.0013}$ & ${0.49}$ & ${-}$ & ${0.007}$ & ${0.020}$ & ${0.04}$ & ${0.04}$ \\

    \cline{2-11}
    & Binary AB 
    & $1654.5/1651$
    
    & $10460.8762$ & $0.0061$ & $79.19$ & $<3.80$ & $3.844$ & $9.853$ & $1.54$ & $19.60$ \\
    
    & & & $0.0056$ & $0.0010$ & $12.19$ & $-$ & $0.009$ & $0.177$ & $0.15$ & $0.03$ \\

    \cline{2-11}
    & Binary BC 
    & $1655.6/1651$
    
    & $10460.8379$ & $0.0157$ & $32.13$ & $<9.77$ & $2.329$ & $8.844$ & $0.73$ & $19.53$ \\
    
    & & & $0.0051$ & $0.0022$ & $4.15$ & $-$ & $0.009$ & $0.307$ & $0.13$ & $0.03$ \\

    \cline{2-11}
    & \textbf{Binary CD}
    & $\boldsymbol{1651.2/1651}$
    
    & $\boldsymbol{10460.8952}$ & $\boldsymbol{0.0169}$ & $\boldsymbol{27.39}$ & $\boldsymbol{<6.46}$ & $\boldsymbol{0.673}$ & $\boldsymbol{8.101}$ & $\boldsymbol{0.57}$ & $\boldsymbol{19.52}$ \\
    
    & & & $\boldsymbol{0.0082}$ & $\boldsymbol{0.0021}$ & $\boldsymbol{3.50}$ & $\boldsymbol{-}$ & $\boldsymbol{0.016}$ & $\boldsymbol{0.388}$ & $\boldsymbol{0.13}$ & $\boldsymbol{0.04}$ \\

    \cline{2-11}
    & Binary DA 
    & $1655.2/1651$
    
    & $10460.8448$ & $0.0040$ & $120.48$ & $<2.51$ & $5.469$ & $9.924$ & $1.91$ & $19.60$ \\
    
    & & & $0.0062$ & $0.0006$ & $17.91$ & $-$ & $0.011$ & $0.171$ & $0.13$ & $0.04$ \\

    
    
    
    \hline
    \hline
    \end{tabular}
    \label{KB241170_parm_2L1Sstatic}
\end{table*}

\begin{figure}
    \includegraphics[width=0.47\textwidth]{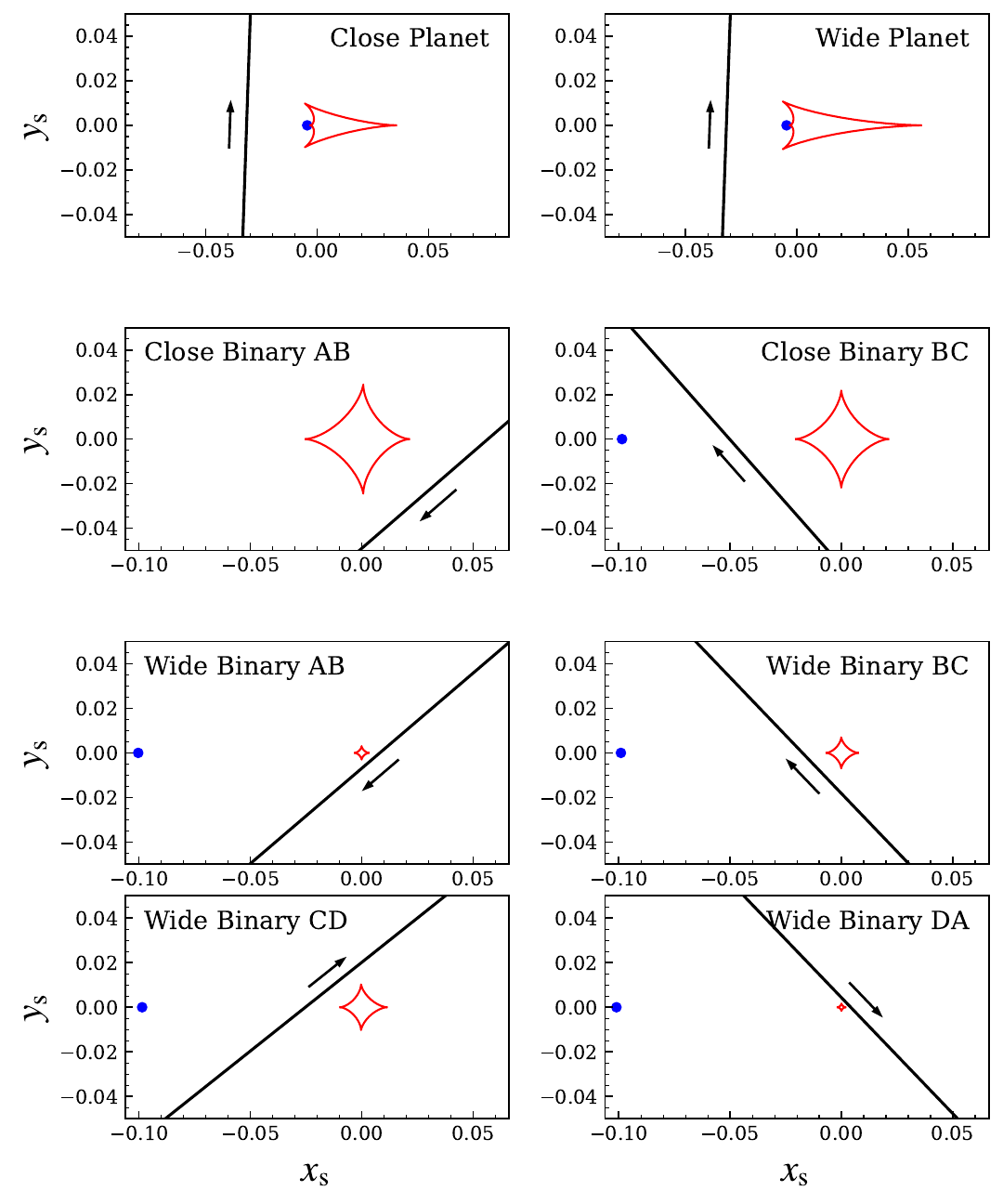}
    \caption{Caustic geometries of the eight 2L1S models for \eventc. The symbols follow the same convention as in Figure~\ref{KB220954_caustic}. Finite-source effects are only marginally detected in all models, so no magenta circles indicating the angular source size are shown.}
\label{KB241170_caustic}
\end{figure}

Figure~\ref{KB241170_lc} displays the observed data around the peak region of \eventc. The $I$-band magnitude has been calibrated to the standard system using the OGLE-III star catalog. The light curve reveals two distinct bumps: the first is sparsely covered by KMTS and KMTC data from $\hjdPrime = 10460.4$ to $10460.7$, while the curvature seen in the KMTC and LCOC data from $\hjdPrime = 10460.75$ to $10460.95$ indicates the interval between the two bumps. However, the second bump itself is not directly covered by observations. 

Similar to the case of \eventb, the grid search yields ten distinct local minima (Figure~\ref{KB241170_gridsearch}), including eight ``Binary'' minima and two ``Planet'' minima. We follow the same notation as in \eventb\ to define the eight non-redundant solutions. Table~\ref{KB241170_parm_2L1Sstatic} lists the 2L1S parameters from the fits, and Figure~\ref{KB241170_caustic} shows the corresponding caustic geometries. Among these, the ``Wide Binary CD'' solution provides the best fit with the lowest $\chi^2$, while the other five ``Binary'' solutions are only modestly disfavored, with $\Delta\chi^2 \leq 4.4$. The two ``Planet'' solutions are more strongly disfavored ($\Delta\chi^2 \sim 18$), but cannot be fully excluded.

For the two ``Planet'' solutions, the mass ratio is $\log q = -2.2$, corresponding to super-Jupiter/Sun mass ratios. The mass ratios of the ``Binary'' solutions span a much broader range, $|\log q| = 0.0$ to $1.9$, reflecting diverse possible interpretations of the binary interpretation. These scenarios will be discussed in more detail in Section~\ref{lens}, in conjunction with the Bayesian analysis. The ``Wide Binary'' solutions also yield longer overall timescales, however, the effective timescale for the body near the source trajectory, given by $\te / \sqrt{1+q}$, is about 13 days, consistent with those of the ``Planet'' and ``Close Binary'' solutions.  

Due to the short event timescale, as with \eventb, inclusion of parallax effects improves the fit only marginally ($\Delta\chi^2 \sim 0.9$), and the parallax vector remains poorly constrained, with $1\sigma$ uncertainties exceeding 3.6 in all directions. We therefore adopt the static models as our final results.

\section{Color-Magnitude Diagram (CMD)}\label{CMD}

\begin{figure*}
    \centering
    \includegraphics[width=0.33\textwidth]{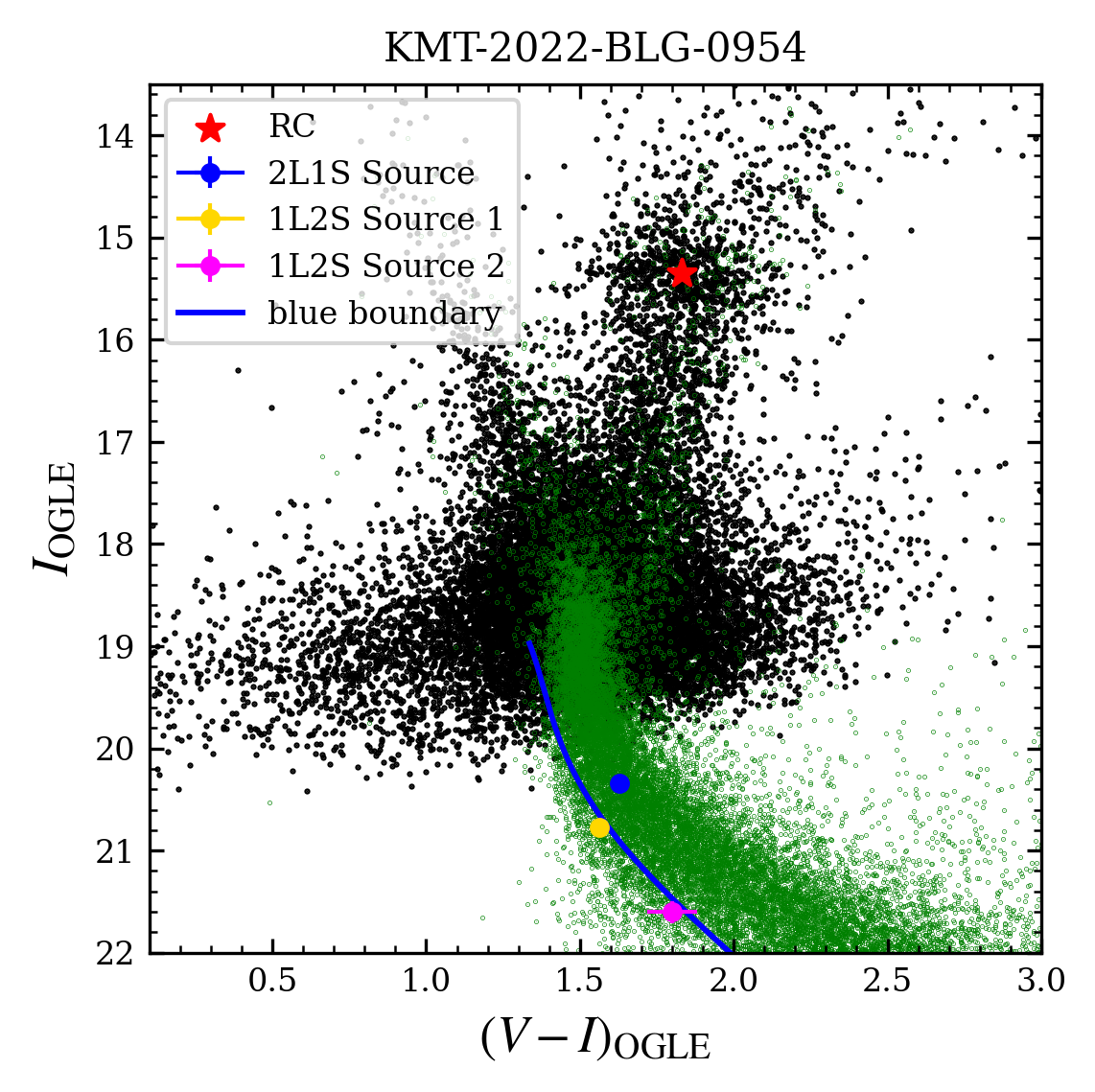}
    \includegraphics[width=0.33\textwidth]{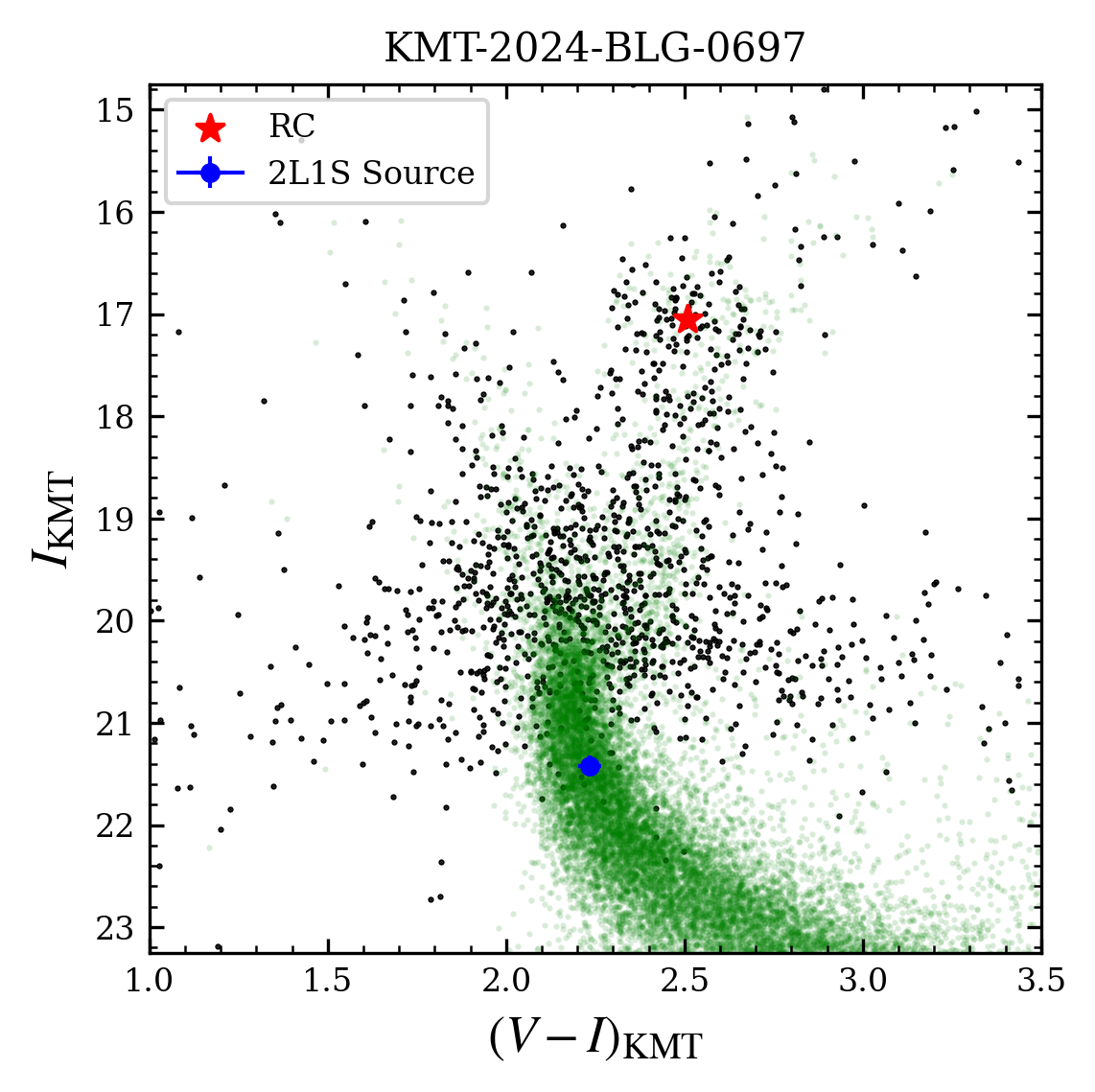}
    \includegraphics[width=0.33\textwidth]{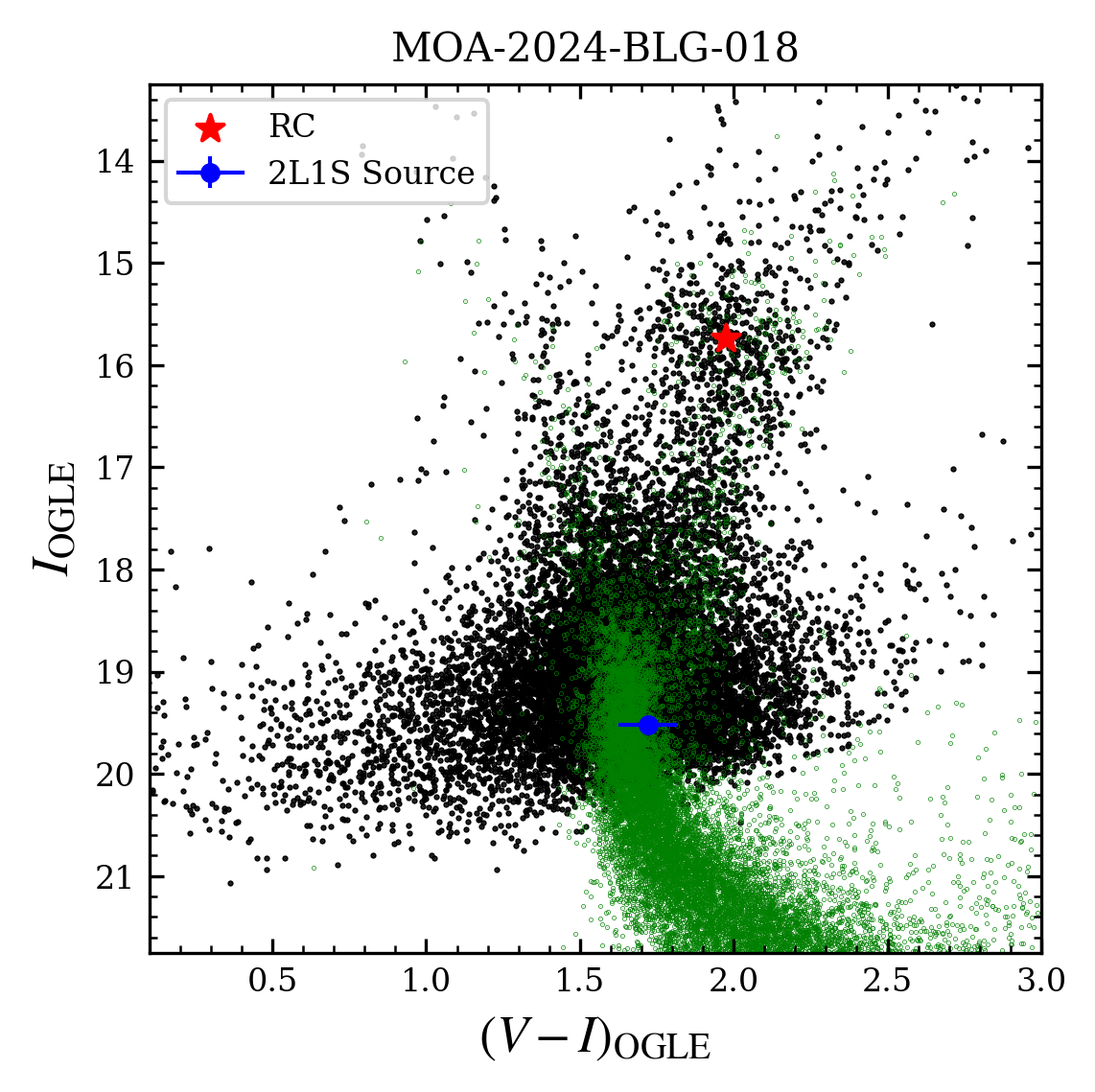}
    \caption{Color-magnitude diagram of the three events. The CMDs (black dots) of \eventa\ and \eventc\ are constructed from the OGLE-III star catalog \citep{OGLEIII} within a $2.5^{\prime}$ radius centered on the event position, while the CMD of \eventb\ is built from KMTC field stars within a $2^\prime \times 2^\prime$ square centered on the event position. The red asterisk marks the centroid of the red clump (RC), and the blue dot indicates the source position for the best-fit 2L1S solution. The green points represent the {\it HST} CMD from \citet{HSTCMD}, whose red clump centroid, $(V - I, I)_{{\rm RC}, HST} = (1.62, 15.15)$ \citep{MB07192}, has been matched to that of KMTC or OGLE-III. 
    For \eventa, the gold and magenta dots represent the primary and secondary sources of the 1L2S model, respectively. The blue line indicates the blue boundary of the bulge main-sequence stars, derived from stellar isochrones with [M/H] = $-1.0$ and age $>9$ Gyr.}
\label{three_cmd}
\end{figure*}

\begin{table*}
    \renewcommand\arraystretch{1.25}
    \centering
    \caption{CMD parameters, 2L1S source properties and derived $\thetae$ and $\murel$ for \eventa}
    
    
    \begin{tabular}{c|c|c  c  r  r r r r}
    \hline
    \hline
    \multicolumn{2}{c|}{Model} & 
    $(V - I, I)_{\rm RC}$ & 
    $(V - I, I)_{\rm RC, 0}$ &
    $(V - I,I)_{\rm S}$ &
    $(V - I,I)_{\rm S,0}$ &
    $\theta_*$ ($\mu$as) &
    $\thetae$ (mas) &
    $\murel$ (${\rm mas\,yr^{-1}}$) \\
    
    \hline
    \multirow{8}{*}{Close} 
    
    & {Planet Finite}
    & $1.830 \pm 0.007$, & $1.06 \pm 0.03$, & $1.627 \pm 0.006$, & $0.86 \pm 0.03$, & $0.49 \pm 0.02$ & $0.078 \pm 0.005$ & $0.64 \pm 0.04$ \\
    
    & & $15.353 \pm 0.020$ & $14.38 \pm 0.04$ & $20.33 \pm 0.02$ & $19.36 \pm 0.05$ &  &  &  \\       

    \cline{2-9}
    & {Planet Point}
    & ↑ & ↑ & $1.627 \pm 0.006$, & $0.86 \pm 0.03$, & $0.49 \pm 0.02$ & $>0.097$ & $>0.78$ \\
    
    & &  &  & $20.35 \pm 0.02$ & $19.38 \pm 0.05$ &  &  &  \\       

    \cline{2-9}
    & Binary AB 
    & ↑ & ↑ & $1.627 \pm 0.006$, & $0.86 \pm 0.03$, & $0.50 \pm 0.02$ & $>0.095$ & $>0.79$ \\
    
    & &  &  & $20.31 \pm 0.02$ & $19.34 \pm 0.05$ &  &  &  \\       

    \cline{2-9}
    & {Binary BC}
    & ↑ & ↑ & $1.627 \pm 0.006$, & $0.86 \pm 0.03$, & $0.49 \pm 0.02$ & $>0.078$ & $>0.63$ \\
    
    & &  &  & $20.34 \pm 0.02$ & $19.37 \pm 0.05$ &  &  &  \\       

    \hline
    \multirow{12}{*}{Wide} 
    & {Planet Finite}
    & ↑ & ↑ & ${1.627 \pm 0.006}$, & ${0.86 \pm 0.03}$, & ${0.49 \pm 0.02}$ & $0.078 \pm 0.005$ & $0.64 \pm 0.04$ \\
    
    & &  &  & ${20.33 \pm 0.02}$ & ${19.36 \pm 0.05}$ &  &  &  \\       

    \cline{2-9}
    & {Planet Point}
    & ↑ & ↑ & $1.627 \pm 0.006$, & $0.86 \pm 0.03$, & $0.49 \pm 0.02$ & $>0.093$ & $>0.74$ \\
    
    & &  &  & $20.35 \pm 0.02$ & $19.38 \pm 0.05$ &  &  &  \\       

    \cline{2-9}
    & Binary AB 
    & ↑ & ↑ & $1.627 \pm 0.006$, & $0.86 \pm 0.03$, & $0.50 \pm 0.02$ & $>0.084$ & $>0.61$ \\
    
    & &  &  & $20.31 \pm 0.02$ & $19.34 \pm 0.05$ &  &  &  \\       

    \cline{2-9}
    & Binary BC 
    & ↑ & ↑ & $1.627 \pm 0.006$, & $0.86 \pm 0.03$, & $0.50 \pm 0.02$ & $>0.262$ & $>0.42$ \\
    
    & &  &  & $20.30 \pm 0.02$ & $19.33 \pm 0.05$ &  &  &  \\       

    \cline{2-9}
    & Binary CD 
    & ↑ & ↑ & $1.627 \pm 0.006$, & $0.86 \pm 0.03$, & $0.50 \pm 0.02$ & $>0.174$ & $>0.47$ \\
    
    & &  &  & $20.30 \pm 0.02$ & $19.33 \pm 0.05$ &  &  &  \\       

    \cline{2-9}
    & {Binary DA}
    & ↑ & ↑ & $1.627 \pm 0.006$, & $0.86 \pm 0.03$, & $0.49 \pm 0.02$ & $>0.093$ & $>0.68$ \\
    
    & &  &  & $20.34 \pm 0.02$ & $19.37 \pm 0.05$ &  &  &  \\       

    \hline
    \hline
    \end{tabular}
    
    \begin{tablenotes}
        \centering
        \item{NOTE. } 
        The parameters are presented with their $1 \sigma$ uncertainties. For the non-detection parameters, the $3 \sigma$ lower limits are provided. 
    \end{tablenotes}
    \label{table_CMD_kb220954}
\end{table*}

\begin{table*}
    \renewcommand\arraystretch{1.25}
    \centering
    \caption{CMD parameters, 2L1S source properties and derived $\thetae$ and $\murel$ for \eventb}
    
    
    \begin{tabular}{c|c|c  c  r  r r r r}
    \hline
    \hline
    \multicolumn{2}{c|}{Model} & 
    $(V - I, I)_{\rm RC}$ & 
    $(V - I, I)_{\rm RC, 0}$ &
    $(V - I,I)_{\rm S}$ &
    $(V - I,I)_{\rm S,0}$ &
    $\theta_*$ ($\mu$as) &
    $\thetae$ (mas) &
    $\murel$ (${\rm mas\,yr^{-1}}$) \\
    
    \hline
    \multirow{4}{*}{Close} 
    
    & {Planet}
    & $2.508 \pm 0.016$, & $1.06 \pm 0.03$, & $2.234 \pm 0.032$, & $0.79 \pm 0.05$, & $0.61 \pm 0.04$ & $0.24 \pm 0.03$ & $5.9 \pm 0.8$ \\
    
    & & $17.050 \pm 0.063$ & $14.35 \pm 0.04$ & $21.48 \pm 0.09$ & $18.78 \pm 0.12$ &  &  &  \\       

    \cline{2-9}
    
    & Binary AB 
    & ↑ & ↑ & $2.234 \pm 0.032$, & $0.79 \pm 0.05$, & $0.62 \pm 0.04$ & $0.28 \pm 0.04$ & $7.1 \pm 1.2$ \\
    
    & &  &  & $21.42 \pm 0.10$ & $18.72 \pm 0.12$ &  &  &  \\

    \hline
    \multirow{4}{*}{Wide} 
    & Planet
    & ↑ & ↑ & $2.234 \pm 0.032$, & $0.79 \pm 0.05$, & $0.60 \pm 0.04$ & $0.24 \pm 0.03$ & $5.8 \pm 0.8$ \\
    
    & &  &  & $21.49 \pm 0.09$ & $18.79 \pm 0.12$ &  &  &  \\

    \cline{2-9}
    & {Binary AB }
    & ↑ & ↑ & ${2.234 \pm 0.032}$, & ${0.79 \pm 0.05}$, & ${0.62 \pm 0.04}$ & ${0.29 \pm 0.04}$ & ${7.1 \pm 1.2}$ \\
    
    & &   &   & ${21.42 \pm 0.10}$ & ${18.72 \pm 0.12}$ &  &  &  \\

    \hline
    \hline
    \end{tabular}
    
    \begin{tablenotes}
        \centering
        \item{NOTE. } 
        The parameters are presented with their $1 \sigma$ uncertainties. 
        $(V - I, I)_{\rm RC}$ and $(V - I,I)_{\rm S}$ are in the instrumental magnitude of KMTC. 
    \end{tablenotes}
    \label{table_CMD_kb240697}
\end{table*}




    
    
    
    
    

\begin{table*}
    \renewcommand\arraystretch{1.25}
    \centering
    \caption{CMD parameters, 2L1S source properties and derived $\thetae$ and $\murel$ for \eventc}
    
    
    \begin{tabular}{c|c|c  c  r  r r r r}
    \hline
    \hline
    \multicolumn{2}{c|}{Model} & 
    $(V - I, I)_{\rm RC}$ & 
    $(V - I, I)_{\rm RC, 0}$ &
    $(V - I,I)_{\rm S}$ &
    $(V - I,I)_{\rm S,0}$ &
    $\theta_*$ ($\mu$as) &
    $\thetae$ (mas) &
    $\murel$ (${\rm mas\,yr^{-1}}$) \\
    
    \hline
    \multirow{6}{*}{Close} 
    
    & {Planet}
    & $1.973 \pm 0.009$, & $1.06 \pm 0.03$, & $1.721 \pm 0.094$, & $0.81 \pm 0.10$, & $0.68 \pm 0.06$ & $>0.041$ & $>1.07$ \\
    
    & & $15.740 \pm 0.038$ & $14.60 \pm 0.04$ & $19.70 \pm 0.04$ & $18.56 \pm 0.07$ &  &  &  \\          

    \cline{2-9}
    & Binary AB 
    & ↑ & ↑ & $1.721 \pm 0.094$, & $0.81 \pm 0.10$, & $0.74 \pm 0.07$ & $>0.052$ & $>1.52$ \\
    
    & &   &   & $19.52 \pm 0.05$ & $18.38 \pm 0.07$ &  &  &  \\        

    \cline{2-9}
    & {Binary BC}
    & ↑ & ↑ & $1.721 \pm 0.094$, & $0.81 \pm 0.10$, & $0.72 \pm 0.07$ & $>0.037$ & $>1.04$ \\
    
    & &   &   & $19.58 \pm 0.05$ & $18.44 \pm 0.07$ &  &  &  \\         

    \hline
    \multirow{10}{*}{Wide} 
    & Planet 
    & ↑ & ↑ & $1.721 \pm 0.094$, & $0.81 \pm 0.10$, & $0.68 \pm 0.06$ & $>0.040$ & $>1.06$ \\
    
    & &   &   & $19.70 \pm 0.04$ & $18.56 \pm 0.07$ &  &  &  \\

    \cline{2-9}
    & Binary AB 
    & ↑ & ↑ & $1.721 \pm 0.094$, & $0.81 \pm 0.10$, & $0.71 \pm 0.07$ & $>0.188$ & $>0.87$ \\
    
    & &  &   & $19.60 \pm 0.03$ & $18.46 \pm 0.06$ &  &  &  \\     

    \cline{2-9}
    & Binary BC 
    & ↑ & ↑ & $1.721 \pm 0.094$, & $0.81 \pm 0.10$, & $0.74 \pm 0.07$ & $>0.075$ & $>0.86$ \\
    
    & &   &   & $19.53 \pm 0.03$ & $18.39 \pm 0.06$ &  &  &  \\    

    \cline{2-9}
    & {Binary CD}
    & ↑ & ↑ & ${1.721 \pm 0.094}$, & ${0.81 \pm 0.10}$, & ${0.74 \pm 0.07}$ & ${>0.115}$ & ${>1.53}$ \\
    
    & &  &  & ${19.52 \pm 0.04}$ & ${18.38 \pm 0.07}$ &  &  &  \\   
    
    \cline{2-9}
    & {Binary DA}
    & ↑ & ↑ & $1.721 \pm 0.094$, & $0.81 \pm 0.10$, & $0.71 \pm 0.07$ & $>0.284$ & $>0.86$ \\
    
    & &   &   & $19.60 \pm 0.04$ & $18.46 \pm 0.07$ &  &  &  \\      

    \hline
    \hline
    \end{tabular}
    
    \begin{tablenotes}
        \centering
        \item{NOTE. } 
        The parameters are presented with their $1 \sigma$ uncertainties. For the non-detection parameters, the $3 \sigma$ lower limits are provided. 
    \end{tablenotes}
    \label{table_CMD_kb241170}
\end{table*}

Before introducing the 1L2S analysis, we perform a CMD analysis in this section for two main reasons.  

First, the CMD analysis provides the angular source radius, $\theta_*$ \citep{Yoo2004}, which can then be used to derive the angular Einstein radius and the lens–source relative proper motion:  
\begin{equation}\label{equ:murel}
    \thetae = \frac{\theta_*}{\rho}, \quad
    \murel = \frac{\thetae}{\te}.
\end{equation}
By studying the $\murel$ distribution of observed planetary microlensing events, \citet{MASADA} and \cite{2019_subprime} showed that the fraction of events with proper motions lower than a given $\murel \ll \sigma_{\mu}$ is  
\begin{equation}\label{eqn:probmu}
p(\leq\murel) \rightarrow \frac{\murel^2}{4\sigma_{\mu}^2} \rightarrow 2.8 \times 10^{-2}\biggl(\frac{\murel}{1\,{\rm mas\,yr^{-1}}}\biggr)^2,
\end{equation}
where we adopt $\sigma_{\mu} = 3.0\,{\rm mas\,yr^{-1}}$ for the proper-motion dispersion of lenses. This provides a way to evaluate a 1L2S model with low $\murel$. 

Second, \cite{Gaudi1998} noted that the 2L1S/1L2S degeneracy can be distinguished by the color difference expected between two sources of different luminosities. Furthermore, \citet{KB220440} showed that the source colors inferred from a 1L2S model may be unphysical (e.g., too blue). The CMD analysis enables us to determine the intrinsic source color and thereby assess the validity of a 1L2S interpretation.

Figure~\ref{three_cmd} shows the $V - I$ versus $I$ CMDs for the three events. For \eventa\ and \eventc, the CMDs are constructed from OGLE-III catalog stars \citep{OGLEIII} located within $2.5^\prime$ of the source position. For \eventb, which lies outside the OGLE-III footprint, the CMD is constructed from stars extracted from KMTC images within a $2^\prime \times 2^\prime$ square centered on the source position. We measure the observed centroid of the red clump (RC), $(V - I, I)_{\rm RC}$, following the method of \citet{Nataf2013}. For the intrinsic centroid of the red clump, $(V - I, I)_{\rm RC,0}$, we adopt $(V - I)_{\rm RC,0} = 1.06 \pm 0.03$, with the value and uncertainty taken from \citet{Bensby2013} and \citet{Nataf2016}, respectively. The value of $I_{\rm RC,0}$ is derived from Table 1 of \citet{Nataf2013}, based on each event’s Galactic coordinates, with an adopted uncertainty of 0.04 magnitude.  

The source color, $(V - I)_{\rm S}$, is determined by regressing $V$-band against $I$-band flux as the lensing magnification varies, using KMTS43 data for \eventa, KMTC19 data for \eventb, and KMTC21 data for \eventc. For \eventa\ and \eventc, the color is further calibrated to the OGLE-III magnitude scale using bright field stars near the event position, matched between KMTNet and OGLE-III. Combining the apparent source magnitude from the light-curve analysis, the intrinsic color and magnitude of the source star are then given by  
\begin{equation}\label{equ:extinction2}
    (V-I, I)_{\rm S,0} = (V-I, I)_{\rm S} - [(V-I, I)_{\rm RC} - (V-I, I)_{\rm RC,0}]. 
\end{equation}

Finally, applying the color–surface brightness relation of \citet{Adams2018},  
\begin{equation}\label{equ:theta*}
    \log(2\theta_*) = 0.378\,(V-I)_{\rm S,0} + 0.542 - 0.2\,I_{\rm S,0}, 
\end{equation}
we obtain the angular source radius, $\theta_*$.  
 
The CMD parameters, 2L1S source properties, and derived $\thetae$ and $\murel$ for the three events are summarized in Tables \ref{table_CMD_kb220954}, \ref{table_CMD_kb240697}, and \ref{table_CMD_kb241170}, respectively.

\section{Single-lens Binary-source Analysis}\label{1L2S}

\begin{table*}
    \renewcommand\arraystretch{1.25}
    \centering
    \caption{1L2S Parameters for the Three Events}
    \begin{tabular}{c|c c c}
    \hline
    \hline
    Event &  \multicolumn{1}{c}{\eventa} & \multicolumn{1}{c}{\eventb} & \multicolumn{1}{c}{\eventc}\\ 
    \hline    
    $\chi^2_{\rm total}$/dof 
    & $9782.3/9795$ & $1135.5/1077$ & $1710.6/1650$\\ 
    $\chi^2_{\rm 1L2S}-\chi^2_{\rm 2L1S, Best}$ 
    & $-2.2$ & $57.5$ & $59.4$\\
    \hline
    
    $t_{0,1}$ (${\rm HJD}^{\prime}$) 
    & $9740.3734 \pm 0.0063$
    & $10425.5987 \pm 0.0022$ 
    & $10460.4727 \pm 0.0073$\\
    
    $t_{0,2}$ (${\rm HJD}^{\prime}$) 
    & $9740.7513 \pm 0.0201$ 
    & $10425.7933 \pm 0.0004$ 
    & $10461.1216 \pm 0.0160$\\
    
    $u_{0,1}$  
    & $0.0121 \pm 0.0004$
    & $0.0105 \pm 0.0012$ 
    & $0.0292 \pm 0.0015$ \\
    
    $u_{0,2}$  
    & $0.0148 \pm 0.0004$
    & $0.0049 \pm 0.0004$ 
    & $0.0402 \pm 0.0026$ \\
    
    $\te$ (days) 
    & $45.61 \pm 0.93$ 
    & $16.36 \pm 1.54$ 
    & $14.74 \pm 0.57$ \\ 
    
    $\rho_{1} (10^{-3})$ 
    & $11.97 \pm 0.46$
    & $10.81 \pm 1.22$ 
    & $32.64 \pm 1.61$\\
    
    $\rho_{2} (10^{-3})$ 
    & $15.55 \pm 0.44$
    & $5.06 \pm 0.47$ 
    & $42.83 \pm 2.41$\\
    
    $q_{f,I}$ 
    & $0.470 \pm 0.066$
    & $1.124 \pm 0.068$ 
    & $1.723 \pm 0.218$ \\  
    
    $q_{f,V}$  
    & $0.380 \pm 0.065$ 
    & $-$ 
    & $-$\\ 
    
    
    $I_{\rm S1}$ 
    & $20.77 \pm 0.05$ 
    & $22.43 \pm 0.13$ 
    & $20.87 \pm 0.07$\\
    
    $I_{\rm S2}$
    & $21.60 \pm 0.11$ 
    & $22.30 \pm 0.10$ 
    & $20.29 \pm 0.09$\\
    $V_{\rm S1}$ 
    & $22.34 \pm 0.05$
    & $-$ 
    & $-$\\
    $V_{\rm S2}$
    & $23.40 \pm 0.14$
    & $-$ 
    & $-$\\
    
    
    
    \hline
    \hline
    \end{tabular}
    \label{parm_1L2S}
\end{table*}

The features observed in the three events, namely the double-bump structures in \eventb\ and \eventc, and the asymmetric peak in \eventa, can also be interpreted within the framework of 1L2S models \citep{Gaudi1998}. Hence, we examine whether such a scenario can adequately reproduce the observed anomalies.  

In a 1L2S configuration, the observed light curve is simply the flux-weighted combination of two independent 1L1S curves. The net magnification at wavelength $\lambda$, $A_{\lambda}(t)$, is given by \citep{MB12486}  
\begin{equation}\label{equ:1L2S_alt}
    A_{\lambda}(t) = \frac{A_{1}(t)F_{{\rm S},1,\lambda} + A_{2}(t)F_{{\rm S},2,\lambda}}
    {F_{{\rm S},1,\lambda} + F_{{\rm S},2,\lambda}}
    = \frac{A_{1}(t) + q_{f,\lambda}A_{2}(t)}{1 + q_{f,\lambda}}, 
\end{equation}
\begin{equation}
    q_{f,\lambda} = \frac{F_{{\rm S},2,\lambda}}{F_{{\rm S},1,\lambda}}, 
\end{equation}
where $F_{{\rm S},j,\lambda}$ denotes the flux of the $j$th source in the bandpass $\lambda$, $A_{j}(t)$ is the corresponding single-lens magnification, and $q_{f,\lambda}$ is the flux ratio of the secondary to the primary source. Here $j=1$ refers to the primary and $j=2$ to the secondary star.  

We search for the 1L2S solutions by employing both MCMC sampling and the Nelder–Mead downhill simplex method. The resulting parameter sets are summarized in Table~\ref{parm_1L2S}. For \eventb\ and \eventc, the 1L2S interpretation is strongly disfavored, with relative $\Delta\chi^2$ values of 57.5 and 59.4 compared to the best-fit 2L1S models. As illustrated by the cumulative $\Delta\chi^2$ curves in Figures~\ref{KB240697_lc} and \ref{KB241170_lc}, these discrepancies originate primarily from the anomalous portions of the light curves and are consistently supported by both KMTNet and LCO datasets. Consequently, the 1L2S models can be rejected for these two events.

For \eventa, the 2L1S and 1L2S models yield comparable $\chi^2$ values, so we include the $V$-band data sets with coverage of the anomaly (KMTC43, KMTS04, and KMTS43). As shown in Table~\ref{parm_1L2S}, the 1L2S model is slightly preferred, with $\Delta\chi^2 = 2.2$ relative to the best-fit 2L1S solution. We therefore examine whether the 1L2S interpretation is physically plausible.  

First, because both sources can independently provide a measurement of $\thetae$, a viable 1L2S solution must satisfy the condition  
\begin{equation}\label{equ:thetae}
    \thetae = \frac{\theta_{*,1}}{\rho_1} = \frac{\theta_{*,2}}{\rho_2}. 
\end{equation}
Based on the CMD analysis in Section~\ref{CMD}, we obtain $\thetae = 0.032 \pm 0.002$ and $0.020 \pm 0.002$ from the primary and secondary sources, respectively, which differ by $4\sigma$. 

Second, the lens–source relative proper motions inferred from the two $\theta_*$ values are $\mu_{\rm rel} = 0.255 \pm 0.017~{\rm mas\,yr^{-1}}$ and $\mu_{\rm rel} = 0.164 \pm 0.017~{\rm mas\,yr^{-1}}$. According to Equation~(\ref{eqn:probmu}), these low values of $\mu_{\rm rel}$ correspond to probabilities an order of magnitude lower than those of the ``Planet Finite'' 2L1S solutions, and about $10^3$ times lower than those of the other 2L1S models that yield only lower limits on $\mu_{\rm rel}$.  

Third, we place the two sources on the CMD with the blue boundary of the bulge main sequence from \citet{KB220440}, which corresponds to a metal-poor population with [M/H] = $-1.0$ and age $>9$ Gyr. See Figure \ref{three_cmd}. Both sources lie outside this boundary, i.e., even bluer than the metal-poor population, by $1.1\sigma$ and $0.6\sigma$, respectively.  

Considering all three tests, we conclude that no physically consistent 1L2S solution can explain the observations. We therefore reject the 1L2S model for \eventa.

\section{Lens Physical Parameters}\label{lens}
\begin{table*}
    \renewcommand\arraystretch{1.8}
    \centering
    \caption{Physical Parameters from Bayesian Analysis for \eventa}

    \begin{tabular}{c|c|cccccc|cc}
    \hline
    \hline
     \multicolumn{2}{c|}{\multirow{3}{*}{Model}}
  & \multicolumn{6}{c|}{Physical Properties} 
  & \multicolumn{2}{c}{$\Delta\chi^2$} \\ 
  
  \multicolumn{2}{c|}{} & $D_{\rm S}$ & $D_{\rm L}$  
  & $M_{1}$ & $M_{2}$  
  & $r_{\perp}$  & $\mu_{\rm hel, rel}$  
  & Gal. Mod. & Light Curve\\ 
  
  \multicolumn{2}{c|}{} & (kpc) & (kpc) 
  & ($M_{\odot}$) &   
  & (au) & (${\rm mas\,yr^{-1}}$) 
  &  & \\ 
  \hline
    
    \multirow{4}{*}{Close} 
    
    & {Planet Finite}
    

    & $8.6^{+0.8}_{-0.7}$ 
    & $7.9^{+0.7}_{-0.7}$ 
    & $0.09^{+0.12}_{-0.04}$ 
    & $6.58^{+9.95}_{-3.43}$ $M_{\oplus}$
    & $0.5^{+0.1}_{-0.1}$ 
    & $0.65^{+0.06}_{-0.06}$ 
    & 4.8 & 0.9 \\ 
    
    \cline{2-10}
    & {Planet Point}
    & $8.8^{+2.2}_{-1.0}$ 
    & $6.5^{+1.2}_{-2.7}$ 
    & $0.61^{+0.35}_{-0.29}$ 
    & $0.78^{+0.52}_{-0.39}$ $M_{\rm Jup}$
    & $1.3^{+0.4}_{-0.4}$ 
    & $3.94^{+2.89}_{-1.49}$ 
    & 0.2 & 17.6 \\ 

    \cline{2-10}
    & Binary AB 
    & $8.9^{+2.1}_{-1.0}$ 
    & $6.6^{+1.2}_{-2.7}$ 
    & $0.50^{+0.29}_{-0.24}$ 
    & $0.10^{+0.09}_{-0.06}$ $M_{\odot}$
    & $0.2^{+0.1}_{-0.1}$ 
    & $4.02^{+2.93}_{-1.50}$ 
    & 0.0 & 19.4 \\ 

    \cline{2-10}
    & {Binary BC}
    & $8.8^{+2.1}_{-1.0}$ 
    & $6.6^{+1.2}_{-2.7}$ 
    & $0.08^{+0.07}_{-0.04}$ 
    & $0.51^{+0.31}_{-0.25}$ $M_{\odot}$
    & $0.3^{+0.1}_{-0.1}$ 
    & $3.89^{+2.88}_{-1.49}$ 
    & 0.0 & 13.3 \\ 

    \hline
    \multirow{6}{*}{Wide} 
    & {Planet Finite}
    
    & $8.6^{+0.8}_{-0.7}$ 
    & $7.9^{+0.7}_{-0.7}$ 
    & $0.09^{+0.12}_{-0.04}$ 
    & $6.49^{+9.15}_{-3.42}$ $M_{\oplus}$
    & $0.8^{+0.1}_{-0.1}$ 
    & $0.65^{+0.06}_{-0.06}$ 
    & 4.8 & 0.0 \\ 

    \cline{2-10}
    & {Planet Point}
    & $8.8^{+2.2}_{-1.0}$ 
    & $6.6^{+1.2}_{-2.7}$ 
    & $0.61^{+0.35}_{-0.29}$ 
    & $0.82^{+0.54}_{-0.40}$ $M_{\rm Jup}$
    & $6.6^{+2.3}_{-2.1}$ 
    & $3.92^{+2.89}_{-1.48}$ 
    & 0.2 & 17.6\\ 

    \cline{2-10}
    & Binary AB 
    & $8.8^{+2.2}_{-1.0}$ 
    & $6.5^{+1.3}_{-2.8}$ 
    & $0.49^{+0.28}_{-0.23}$ 
    & $0.12^{+0.11}_{-0.07}$ $M_{\odot}$
    & $44.6^{+18.4}_{-15.1}$ 
    & $3.67^{+2.87}_{-1.43}$ 
    & 0.7 & 18.6\\ 

    \cline{2-10}
    & Binary BC 
    & $8.4^{+1.5}_{-1.0}$ 
    & $3.7^{+3.2}_{-1.6}$ 
    & $0.03^{+0.01}_{-0.01}$ 
    & $0.77^{+0.28}_{-0.31}$ $M_{\odot}$
    & $90.6^{+22.5}_{-22.7}$ 
    & $1.53^{+1.74}_{-0.76}$ 
    & 12.2 & 22.9\\ 

    \cline{2-10}
    & Binary CD 
    & $8.6^{+1.9}_{-1.1}$ 
    & $4.4^{+3.0}_{-2.1}$ 
    & $0.08^{+0.04}_{-0.04}$ 
    & $0.66^{+0.29}_{-0.28}$ $M_{\odot}$
    & $87.6^{+25.0}_{-24.2}$ 
    & $2.07^{+2.06}_{-0.95}$ 
    & 7.8 & 24.3 \\ 

    \cline{2-10}
    & {Binary DA}
    & $8.8^{+2.2}_{-1.0}$ 
    & $6.5^{+1.3}_{-2.8}$ 
    & $0.52^{+0.29}_{-0.24}$ 
    & $0.10^{+0.08}_{-0.06}$ $M_{\odot}$
    & $40.1^{+16.3}_{-13.4}$ 
    & $3.71^{+2.88}_{-1.43}$ 
    & 0.7 & 13.3\\ 

    \hline
    \hline
    \end{tabular}
    
    \begin{tablenotes}
        \centering
        \item{NOTE. } 
        The parameters are presented with their $1 \sigma$ uncertainties. The $\Delta\chi^2$ of Gal.Mod. is derived by $-2\ln({ P_{\rm Gal.Mod.}})$, where $P_{\rm Gal.Mod.}$ represents the relative probability from the Galactic model. The $\Delta\chi^2$ of light-curve analysis is from Table \ref{KB220954_parm_2L1Sstatic}. 
        
    \end{tablenotes}
    \label{table_bayes_kb220954}
\end{table*}

\begin{table*}
    \renewcommand\arraystretch{1.8}
    \centering
    \caption{Physical Parameters from Bayesian Analysis for \eventb}
   
    \begin{tabular}{c|c|cccccc|cc}
    \hline
    \hline
     \multicolumn{2}{c|}{\multirow{3}{*}{Model}}
  & \multicolumn{6}{c|}{Physical Properties} 
  & \multicolumn{2}{c}{$\Delta\chi^2$} \\ 
  
  \multicolumn{2}{c|}{} & $D_{\rm S}$ & $D_{\rm L}$  
  & $M_{1}$ & $M_{2}$  
  & $r_{\perp}$  & $\mu_{\rm hel, rel}$  
  & Gal. Mod. & Light Curve \\ 
  
  \multicolumn{2}{c|}{} & (kpc) & (kpc) 
  & ($M_{\odot}$) &  ($M_{\rm Jup}$ ) 
  & (au) & (${\rm mas\,yr^{-1}}$) 
  &  &  \\ 
  \hline
    
    \multirow{2}{*}{Close} 
    
    & {Planet}
    & $8.5^{+1.3}_{-0.8}$ 
    & $7.1^{+0.7}_{-0.8}$ 
    & $0.31^{+0.27}_{-0.16}$ 
    & $0.77^{+0.68}_{-0.40}$ 
    & $1.1^{+0.2}_{-0.2}$ 
    & $5.97^{+0.90}_{-0.83}$ 
    & 0.0 & 8.0 \\ 

    \cline{2-10}
    & Binary AB 
    & $8.5^{+1.2}_{-0.8}$
    & $7.0^{+0.7}_{-0.9}$
    & $0.35^{+0.27}_{-0.18}$  
    & $17.39^{+14.86}_{-9.22}$ 
    & $0.4^{+0.1}_{-0.1}$
    & $7.08^{+1.17}_{-1.07}$
    & 0.0 & 0.5 \\ 

    \hline
    \multirow{2}{*}{Wide} 
    & {Planet}
    & $8.5^{+1.3}_{-0.8}$ 
    & $7.2^{+0.7}_{-0.8}$ 
    & $0.31^{+0.27}_{-0.16}$  
    & $0.77^{+0.68}_{-0.40}$ 
    & $2.7^{+0.4}_{-0.4}$ 
    & $5.90^{+0.89}_{-0.82}$ 
    & 0.1  & 7.9 \\  

    \cline{2-10}
    & Binary AB 
    & $8.5^{+1.3}_{-0.8}$ 
    & $7.0^{+0.7}_{-0.9}$ 
    & $0.35^{+0.27}_{-0.18}$  
    & $18.92^{+16.51}_{-10.12}$ 
    & $9.8^{+1.9}_{-1.9}$ 
    & $6.92^{+1.14}_{-1.05}$ 
    & 0.1 & 0.0 \\  

    \hline
    \hline
    \end{tabular}
    \label{table_bayes_kb240697}
\end{table*}

\begin{table*}
    \renewcommand\arraystretch{1.8}
    \centering
    \caption{Physical Parameters from Bayesian Analysis for \eventc}
   
    \begin{tabular}{c|c|cccccc|cc}
    \hline
    \hline
     \multicolumn{2}{c|}{\multirow{3}{*}{Model}}
  & \multicolumn{6}{c|}{Physical Properties} 
  & \multicolumn{2}{c}{$\Delta\chi^2$} \\ 
  
  \multicolumn{2}{c|}{} & $D_{\rm S}$ & $D_{\rm L}$  
  & $M_{1}$ & $M_{2}$  
  & $r_{\perp}$  & $\mu_{\rm hel, rel}$  
  & Gal. Mod. & Light Curve \\ 
  
  \multicolumn{2}{c|}{} & (kpc) & (kpc) 
  &  &   
  & (au) & (${\rm mas\,yr^{-1}}$) 
  &  &\\ 
  \hline
    
    \multirow{3}{*}{Close} 
    
    & {Planet}
    & $9.7^{+0.9}_{-1.0}$ 
    & $8.0^{+0.9}_{-1.5}$ 
    & $0.30^{+0.32}_{-0.19}$ $M_{\odot}$
    & $1.86^{+2.00}_{-1.16}$ $M_{\rm Jup}$ 
    & $1.2^{+0.5}_{-0.4}$ 
    & $6.16^{+2.67}_{-2.20}$ 
    & 0.2 & 17.5 \\

    \cline{2-10}
    & Binary AB 
    & $9.7^{+0.9}_{-1.0}$ 
    & $8.0^{+0.9}_{-1.4}$ 
    & $0.11^{+0.12}_{-0.07}$ $M_{\odot}$
    & $0.17^{+0.19}_{-0.11}$ $M_{\odot}$
    & $0.4^{+0.2}_{-0.1}$ 
    & $6.34^{+2.70}_{-2.23}$ 
    & 0.0 & 0.2 \\

    \cline{2-10}
    & Binary BC
    & $9.7^{+0.9}_{-1.0}$ 
    & $8.0^{+0.9}_{-1.4}$ 
    & $0.15^{+0.17}_{-0.09}$ $M_{\odot}$
    & $0.13^{+0.15}_{-0.08}$ $M_{\odot}$
    & $0.3^{+0.1}_{-0.1}$ 
    & $6.28^{+2.68}_{-2.22}$ 
    & 0.1 & 3.6\\

    \hline
    \multirow{5}{*}{Wide} 
    & {Planet}
    & $9.7^{+0.9}_{-1.0}$ 
    & $8.0^{+0.9}_{-1.5}$ 
    & $0.30^{+0.32}_{-0.19}$ $M_{\odot}$
    & $1.91^{+2.05}_{-1.19}$ $M_{\rm Jup}$ 
    & $2.5^{+1.0}_{-0.9}$ 
    & $6.16^{+2.66}_{-2.20}$ 
    & 0.2 & 17.9 \\ 

    \cline{2-10}
    & Binary AB 
    & $9.5^{+1.1}_{-1.3}$ 
    & $4.9^{+3.1}_{-2.4}$ 
    
    & $18.06^{+14.98}_{-9.62}$ $M_{\rm Jup}$ 
    
    & $0.58^{+0.37}_{-0.29}$ $M_{\odot}$
    & $29.1^{+10.4}_{-9.2}$ 
    & $3.22^{+2.60}_{-1.48}$ 
    & 8.3 & 3.3\\ 

    \cline{2-10}
    & Binary BC 
    & $9.7^{+1.0}_{-1.1}$ 
    & $7.3^{+1.3}_{-2.8}$ 
    & $0.08^{+0.07}_{-0.05}$ $M_{\odot}$
    & $0.39^{+0.32}_{-0.23}$ $M_{\odot}$
    & $22.0^{+8.5}_{-8.1}$ 
    & $4.55^{+2.77}_{-1.79}$ 
    & 3.3 & 4.4 \\ 

    \cline{2-10}
    & Binary CD 
    & $9.7^{+1.0}_{-1.1}$ 
    & $7.5^{+1.1}_{-2.6}$ 
    & $0.10^{+0.09}_{-0.06}$ $M_{\odot}$
    & $0.35^{+0.29}_{-0.20}$ $M_{\odot}$
    & $19.3^{+7.4}_{-6.6}$ 
    & $4.91^{+2.67}_{-1.78}$ 
    & 2.6 & 0.0\\ 

    \cline{2-10}
    & Binary DA 
    & $9.4^{+1.2}_{-1.3}$ 
    & $4.0^{+3.2}_{-1.9}$ 
    
    & $8.70^{+6.07}_{-4.39}$ $M_{\rm Jup}$ 
    
    & $0.67^{+0.36}_{-0.32}$ $M_{\odot}$
    & $31.4^{+9.6}_{-9.6}$ 
    & $2.68^{+2.16}_{-1.21}$ 
    & 11.3 & 4.0 \\ 

    \hline
    \hline
    \end{tabular}
    \label{table_bayes_kb241170}
\end{table*}

By combining Equations~(\ref{eqn:te}) and (\ref{equ:pie}), the lens mass $M_{\rm L}$ and distance $D_{\rm L}$ can be expressed in terms of the angular Einstein radius and the microlensing parallax \citep{Gould1992, Gould2000}:
\begin{equation}\label{eq:mass}
M_{\rm L} = \frac{\thetae}{\kappa \pie}, \qquad
D_{\rm L} = \frac{{\rm au}}{\pie\thetae + \pi_{\rm S}} .
\end{equation}
Because all surviving light-curve solutions yield relatively large uncertainties in $\pie$, and some do not provide unambiguous determinations of $\thetae$, we estimate the lens properties through a Bayesian analysis that incorporates priors from a Galactic population model.

The adopted Galactic model is based on three key components: the stellar mass function, the spatial density profile of stars, and their kinematic distributions. For the total lens mass, we follow the mass function of \citet{Kroupa2001}, with applying an upper cutoff of $1.3M_{\odot}$ for disk lenses and $1.1M_{\odot}$ for bulge lenses \citep{Zhu2017spitzer}. The stellar density profiles are taken from the models of \citet{Yang2021_GalacticModel}. For the kinematics, we adopt the dynamically self-consistent ``Model C” of \citet{Yang2021_GalacticModel} for the Galactic disk, and for the bulge we use the prescription of \citet{Zhu2017spitzer}, assuming zero mean velocity and a velocity dispersion of $120~{\rm km~s^{-1}}$ in each direction.

For each case, we simulate a sample of $5 \times 10^7$ events. Each simulated event $i$, with parameters $\tEi$, $\mu_{{\rm rel},i}$, and $\theta_{{\rm E},i}$, is assigned a weight
\begin{equation}\label{equ: weight}
w_{i} = \Gamma_{i}\times p(\tEi), p(\theta_{{\rm E},i}),
\end{equation}
where $\Gamma_{i} = \theta_{{\rm E},i} \times \mu_{{\rm rel},i}$ represents the microlensing event rate, $p(\tEi)$ is the probability of $\tEi$ given the MCMC error distributions, and $p(\theta_{{\rm E},i})$ is the probability of $\theta_{{\rm E},i}$. To evaluate $p(\theta_{{\rm E},i})$, we first calculate $\rho_i = \theta_* / \theta_{{\rm E},i}$ and then determine the corresponding $\chi^2(\rho_i)$ from the lower envelope of the $\chi^2$–$\rho$ diagram. For \eventa, because there are meaningful constraints on $\pie$, we further weight the simulated events by the posterior distribution of $\pie$ from MCMC. The two lens bodies, are derived by $M_1 = M_{\rm L}/(1+q)$ and $M_2 = M_{\rm L}q/(1+q)$, respectively.

\subsection{\eventa}

The physical parameters derived from the Bayesian analysis for \eventa\ are summarized in Table~\ref{table_bayes_kb220954}, including the component lens masses, the source and lens distances, the projected lens–lens separations, and the heliocentric lens–source relative proper motion, $\mu_{\rm hel, rel}$. Table~\ref{table_bayes_kb220954} also shows the $\Delta\chi^2$ between different models based on the relative probability from the Galactic model and the light-curve analysis. 

For the four ``Planet'' solutions, the planet lies either near or well beyond the snow line, adopting a snow-line radius of $a_{\rm SL} = 2.7(M/M_{\odot})~{\rm au}$ \citep{snowline}. However, the inferred lens masses differ substantially between the ``Planet Finite'' and ``Planet Point'' solutions. In the two ``Planet Finite'' cases, the very small Einstein radius of $\thetae = 0.078 \pm 0.005$ mas leads the Bayesian analysis to favor a super-Earth orbiting a late-type M dwarf or possibly a brown dwarf. In contrast, the ``Planet Point'' solutions suggest a Jupiter-mass planet orbiting a K- or M-dwarf host. The two ``Close Binary'' solutions are consistent with a pair of M dwarfs separated by 0.1–0.4 au. The ``Wide Binary AB'' and ``Wide Binary DA'' solutions also point to M-dwarf pairs, while the ``Wide Binary BC'' and ``Wide Binary CD'' solutions favor a K-dwarf primary with a low-mass M-dwarf or brown-dwarf companion.

The ``Planet Finite'' solutions yield the lowest $\chi^2$ in the light-curve analysis, whereas the ``Close Binary'' and ``Planet Point'' solutions are preferred in the Bayesian analysis. In contrast, the ``Wide Binary BC'' and ``Wide Binary CD'' solutions have the lowest relative probabilities because their very long timescales are disfavored under the Galactic model.


\subsection{\eventb}

The posterior distributions of the physical parameters for \eventb\ are shown in Table~\ref{table_bayes_kb240697}. The two ``Planet'' solutions most likely correspond to a Jupiter-mass planet orbiting an M dwarf beyond the snow line. In contrast, the two ``Binary'' solutions suggest a lens system composed of an M dwarf and a low-mass brown dwarf. All four solutions have nearly the same relative probabilities under the Galactic model. The lens system is likely located in the Galactic bulge.

\subsection{\eventc}

Table~\ref{table_bayes_kb241170} presents the physical parameters for \eventc. The two ``Planet'' solutions correspond to a Jupiter-class planet orbiting an M dwarf. The ``Close Binary'' solutions consist of a pair of low-mass M dwarfs. These four solutions have nearly equal relative probabilities under the Galactic model. The ``Wide Binary BC'' and ``Wide Binary CD'' solutions favor an M-dwarf primary with a low-mass M-dwarf or brown-dwarf companion. The ``Wide Binary AB'' and ``Wide Binary DA'' solutions instead suggest a K/M-dwarf host with a low-mass brown-dwarf or super-Jupiter companion. Similar to \eventa, the`` Wide Binary AB'' and ``Wide Binary DA'' solutions are assigned the lowest relative probabilities by the Galactic model due to their long timescales.

\section{Discussion}\label{dis}
\begin{table*}
    \renewcommand\arraystretch{1.5}
    \centering
    \caption{Known Events Subject to the ``Planet/Binary'' Degeneracy}
    \label{historyevents}
    
    \begin{threeparttable}
    \begin{tabular}{l l l l}
    \hline
    \hline
    Event Name & Reference & $\chi^2_{\rm Binary, best}$ - $\chi^2_{\rm Planet, best}$ & $\murel$ Information from the Light-curve Analysis \\
    \hline 
    OGLE-2011-BLG-0526 & \citet{OB110950} & 2.9 & unmeasurable for Planet model\tnote{1} \\
    \hline
    
    OGLE-2011-BLG-0950 & \citet{OB110950} &  & \\
    & \citet{Sean2022} & 26.7 & $\murel$ = $1.05 \pm 0.20$ ${\rm mas\,yr^{-1}}$ for Planet model  \\
    & & & unmeasurable for Binary model\\
    \hline   
    OGLE-2012-BLG-0455 & \citet{Park2014} & 5.1 
    & $\murel$ = $2.91 \pm 0.27$ ${\rm mas\,yr^{-1}}$ for Binary model\\
    &&& $\murel$ = $3.68 \pm 0.30$ ${\rm mas\,yr^{-1}}$ for Planet model \\
    \hline
    
    
    MOA-2015-BLG-337   & \citet{Miyazaki2018} & 6.1 
    &   $\murel$ > $2.26$ ${\rm mas\,yr^{-1}}$ for Close Binary model\tnote{2}\\
    &&& $\murel$ > $2.11$ ${\rm mas\,yr^{-1}}$ for Wide Binary model\tnote{2}\\
    &&& $\murel$ = $1.90\pm0.29$ ${\rm mas\,yr^{-1}}$ for Close Planet model\\
    &&& $\murel$ = $1.59\pm0.18$ ${\rm mas\,yr^{-1}}$ for Wide Planet model\\
    \hline
    
    OGLE-2016-BLG-1704 & \citet{Paper11Shin2024} & -2.0 & unmeasurable for all models\\
    \hline
    
    KMT-2017-BLG-0958  & \citet{2017_subprime} & -5.5 & unmeasurable for all models\\
    \hline
    
    KMT-2018-BLG-2164  & \citet{Paper5Gould2022} & 4.7 & unmeasurable for all models\\
    \hline
    
    OGLE-2018-BLG-1554 & \citet{Paper5Gould2022} & 0.0 & $\murel$ = $0.73\pm0.20$ ${\rm mas\,yr^{-1}}$ for Close Planet model\\
    &&& $\murel$ = $1.01\pm0.51$ ${\rm mas\,yr^{-1}}$ for Wide Planet model\\
    &&&unmeasurable for Binary model\\
    \hline
    
    KMT-2018-BLG-2718  & \citet{Paper5Gould2022} & 12.7 & unmeasurable for all models\\
    \hline
    
    {\eventa} & this work & 13.3 & $\murel$ = $0.64\pm0.04$ ${\rm mas\,yr^{-1}}$ for Planet Finite model\\
    &&& unmeasurable for Planet Point and Binary models\\
    \hline
    
    KMT-2023-BLG-1896  & \citet{KB231896} & 9.9 & unmeasurable for all models\\
    \hline
   
    {\eventb} & this work & -7.9 
    & $\murel$ = $7.1 \pm 1.2$ ${\rm mas\,yr^{-1}}$ for Binary model\\
    &&& $\murel$ = $5.8 \pm 0.8$ ${\rm mas\,yr^{-1}}$ for Planet model \\
    \hline    
    {\eventc} & this work & -17.5 & unmeasurable for all models\\  

    \hline
    \hline
    \end{tabular}
    \begin{tablenotes}
    \item[1] The proper motions for the binary models were not provided by \cite{OB110950}.
    \item[2] The $1 \sigma$ lower limit is provided. 
    \end{tablenotes}
    
    
    \end{threeparttable}
   
\end{table*}

\subsection{Resolving the ``Planet/Binary'' Degeneracy}

In this paper, we have analyzed three events that exhibit the ``Planet/Binary'' degeneracy, as part of a follow-up program for KMTNet HM events. Prior to this work, 12 unambiguous planetary events from this program had been published, suggesting that the ``Planet/Binary'' degeneracy could represent a non-negligible fraction of the final sample. In addition, another event with this degeneracy, KMT-2025-BLG-1314 (Ren et al., in preparation), has been identified, implying that the occurrence rate of the ``Planet/Binary'' degeneracy may be even higher than that of the KMTNet planetary sample \citep{OB160007}. This outcome is not unexpected, because such degeneracies naturally arise in HM events. 

Including the three events analyzed in this work, there are now 13 known cases subject to the ``Planet/Binary'' degeneracy. Table~\ref{historyevents} summarizes the name and $\Delta\chi^2$ values between the best-fit binary and planetary models. For consistency, we select events with $|\Delta\chi^2|<36$ (i.e., within $6\sigma$) between the binary and planetary solutions. 

The ``Planet/Binary'' degeneracy can, in some cases, be resolved through follow-up high-resolution imaging. For OGLE-2011-BLG-0950, Keck adaptive optics (AO) observations resolved the lens from the source and measured $\mu_{\rm rel} = 4.06 \pm 0.22~{\rm mas,yr^{-1}}$, which is inconsistent with the $\mu_{\rm rel} = 1.05 \pm 0.20~{\rm mas,yr^{-1}}$ predicted by the planetary solutions from the light-curve analysis \citep{Sean2022}. Because finite-source effects are only marginally constrained for the binary models, the planetary solutions are ruled out, leaving the binary interpretation as viable. Table \ref{historyevents} also lists the inferred $\mu_{\rm rel}$ from the light-curve analysis for different models. Similar resolutions could be achieved for two other cases, OGLE-2018-BLG-1554 \citep{Paper5Gould2022} and MOA-2015-BLG-337 \citep{Miyazaki2018}. In both events, $\mu_{\rm rel}$ is well measured for the planetary solutions, while the binary solutions predict only a lower limit. Thus, high-resolution imaging could be used to exclude the planetary interpretations. For OGLE-2011-BLG-0526, the situation is reversed: the binary solutions yield a determination of $\mu_{\rm rel}$, whereas the planetary solutions do not. In this case, high-resolution imaging could potentially reject the binary solutions. 

However, high-resolution imaging is unlikely to break the ``Planet/Binary'' degeneracy for the other nine cases. In six of these, the finite-source effects are only marginally constrained in all models. For two events, OGLE-2012-BLG-0455 \citep{Park2014} and \eventb, the predicted $\mu_{\rm rel}$ values differ by $\lesssim 2\sigma$ between the planetary and binary solutions, making it difficult for high-resolution imaging to provide a decisive resolution. For \eventa, although the ``Planet Finite'' solutions yield a determination of $\mu_{\rm rel}$ that could be tested and potentially rejected by high-resolution imaging, the ``Planet Point'' and binary solutions predict only lower limits, so the ``Planet/Binary'' degeneracy cannot be fully resolved using the classical high-resolution imaging method \citep{Sean2022}.

Besides light-curve modeling and high-resolution imaging, another potential approach to breaking the ``Planet/Binary'' degeneracy is satellite parallax, i.e., observing the same microlensing event simultaneously from Earth and one or more well-separated satellites \citep{1966MNRAS.134..315R,1994ApJ...421L..75G,Gould1995single}. Because the lens–source relative trajectories differ between widely separated observatories, planetary and binary solutions can produce distinct light curves as seen from one location, even if the degeneracy persists at another. This prospect has been demonstrated with the \textit{Spitzer} satellite. For OGLE-2015-BLG-1212, the inclusion of \textit{Spitzer} data increased the exclusion significance of the planetary solutions from $\Delta\chi^2 = 53$ to $114$ \citep{Bozza2016}.

The upcoming space-based microlensing surveys, the Roman Galactic Bulge Time-Domain Survey \citep{MatthewWFIRSTI} and the Earth 2.0 microlensing survey \citep{ET}, will be carried out by satellites located at the halo orbit of the Sun–Earth L2 point. The projected Earth–satellite separation is about two orders of magnitude shorter than that of \textit{Spitzer} ($\sim0.01~\rm au$ versus $\sim1~\rm au$). On the one hand, this shorter baseline may make the paired light curves more sensitive to the differences in the central caustics of the planetary and binary solutions. On the other hand, it remains unclear whether a $\sim0.01~\rm au$ separation is sufficient to reliably capture these differences. We will investigate this question further in a forthcoming study (Zhang et al., in preparation).

\subsection{A New Degeneracy}

The degeneracy between the ``Planet Finite'' and ``Planet Point'' solutions for \eventa\ represents a previously unrecognized type of degeneracy, fundamentally different from the classical $(\rho,q)$ continuous degeneracy associated with the planetary caustic described by \citet{Gaudi&Gould1997ApJ_continuous_degeneracy}. Unlike that case, the new degeneracy is discrete in nature, appearing as distinct minima in parameter space rather than a continuous valley. Similar to three other events in which $\mu_{\rm rel}$ is determined only for the planetary solutions, the source size in the ``Planet Finite'' solution is larger than the central caustic, which yields a low $\mu_{\rm rel}$ that can be tested by high-resolution imaging. However, because the ``Planet Point'' and binary solutions have similarly weak constraints on $\mu_{\rm rel}$, the ``Planet/Binary'' degeneracy cannot be fully resolved by high-resolution imaging alone. Interestingly, this new degeneracy is also found in KMT-2025-BLG-1314 (Ren et al., in preparation), suggesting that it may be common in comparable events. It is therefore worthwhile to re-examine published cases and carefully search for this degeneracy in future events, as its presence could affect microlensing planetary statistics.

\section*{Acknowledgements}
J.Z., W.Zang, S.M., H.Y., Y.T., Q.Q., and W.Zhu acknowledge support by the National Natural Science Foundation of China (Grant No. 12133005). W.Zang acknowledges the support from the Harvard-Smithsonian Center for Astrophysics through the CfA Fellowship. This research has made use of the KMTNet system operated by the Korea Astronomy and Space Science Institute (KASI) and the data were obtained at three host sites of CTIO in Chile, SAAO in South Africa, and SSO in Australia. This research uses data obtained through the Telescope Access Program (TAP), which has been funded by the TAP member institutes. This research was supported by the Korea Astronomy and Space Science Institute under the R\&D program (Project No. 2025-1-830-5) supervised by the Ministry of Science and ICT. This work makes use of observations from the Las Cumbres Observatory global telescope network. The MOA project is supported by JSPS KAKENHI Grant Number JSPS24253004, JSPS26247023, JSPS23340064, JSPS15H00781, JP16H06287, and JP17H02871. H.Y. acknowledges support by the China Postdoctoral Science Foundation (No. 2024M762938). The OGLE project has received funding from the Polish National Science Centre grant OPUS-28 2024/55/B/ST9/00447 to AU. Work by C.H. was supported by the grants of National Research Foundation of Korea (2019R1A2C2085965 and 2020R1A4A2002885). Y.S. acknowledges support from BSF Grant No. 2020740. Work by J.C.Y. acknowledges support from N.S.F Grant No. AST-2108414. W.Zhu acknowledges the science research grants from the China Manned Space Project with No.\ CMS-CSST-2021-A11. The authors acknowledge the Tsinghua Astrophysics High-Performance Computing platform at Tsinghua University for providing computational and data storage resources that have contributed to the research results reported within this paper.  

\section*{Data Availability}
Data used in the light-curve analysis are provided along with publication as supplementary data.



\bibliographystyle{mnras}
\bibliography{Zang.bib} 

\bsp	
\label{lastpage}
\end{document}